\documentclass[twocolumn]{aastex62}
\newcommand{\hi}{H\,\textsc{i}}
\newcommand{\kms}{km\,s$^{-1}$}
\newcommand{\Jykms}{Jy\,\kms}
\newcommand{\Msol}{M$_\odot$}
\newcommand{\HIit}{\mbox{H\hspace{0.155 em}{\footnotesize \it I}}}
 \shorttitle{WAVES I}
 \shortauthors{Minchin et al.}
 \begin{document}
 \title{The Widefield Arecibo Virgo Extragalactic Survey I: New structures in the ALFALFA Virgo 7 cloud complex and an extended tail on NGC 4522}
 \author[0000-0002-1261-6641]{Robert F. Minchin}
 \affiliation{Stratospheric Observatory for Infrared Astronomy/USRA, NASA Ames Research Center, MS 232-12, Moffett Field, CA 94035, USA}
 \author{Rhys Taylor}
 \affiliation{Astronomical Institute of the Czech Academy of Sciences, Bocni II 1401/1a, 141 31 Praha 4, Czech Republic}
 \author{Joachim K\"oppen}
 \affiliation{Institut f\"ur Theoretische Physik und Astrophysik der Universit\"at zu Kiel, D-24098, Kiel, Germany}
 \author{Jonathan I. Davies}
 \affiliation{School of Physics and Astronomy, Cardiff University, Queen's Buildings, The Parade, Cardiff, CF24 3AA, United Kingdom}
 \author{Wim van Driel}
 \affiliation{GEPI, Observatoire de Paris, PSL Research University, CNRS, 5 place Jules Janssen, F-92190 Meudon, France; Station de Radioastronomie de Nan\c{c}ay, Observatoire de Paris, CNRS/INSU USR 704, Université d'Orl\'{e}ans OSUC, route de Souesmes, F-18330 Nan\c{c}ay, France}
 \author{Olivia Keenan}
 \affiliation{School of Physics and Astronomy, Queen Mary, University of London, G O Jones Building, 329 Mile End Rd, London, E1 4NT, United Kingdom}
 \correspondingauthor{Robert Minchin}
 \email{rminchin@usra.edu}

 \begin{abstract}
 We are carrying out a sensitive blind survey for neutral hydrogen (\hi) in the Virgo cluster and report here on the first  $5\degr \times 1\degr$ area covered, which includes two optically-dark gas features: the five-cloud ALFALFA Virgo 7 complex \citep{Kent2007,Kent2009} and the stripped tail of NGC 4522 \citep{Kenney2004}. We discover a sixth cloud and low velocity gas that extends the velocity range of the complex to over 450 \kms, find that around half of the total \hi\ flux comes from extended emission rather than compact clouds, and see around 150 percent more gas, raising the total \hi\ mass from $5.1 \times 10^8$ \Msol\ to $1.3 \times 10^9$ \Msol. This makes the identification of NGC 4445 and NGC 4424 by \citet{Kent2009} as possible progenitors of the complex less likely, as it would require an unusually high fraction of the gas removed to have been preserved in the complex. We also identify a new component to the gas tail of NGC 4522 extending to $\sim 200$ \kms\ below the velocity range of the gas in the galaxy, pointing towards the eastern end of the complex. We consider the possibility that NGC 4522 may be the parent galaxy of the complex, but the large velocity separation ($\sim 1800$ \kms) leads us to rule this out. We conclude that, in the absence of any better candidate, NGC 4445 remains the most likely parent galaxy, although this requires it to have been particularly gas-rich prior to the event that removed its gas into the complex.
 \end{abstract}
 
 \section{Introduction}
 
 Virgo has long been a fertile ground for searches for neutral hydrogen (\hi) gas unassociated with optical counterparts. These include well known examples such as HI 1225+01 \citep{Giovanelli1989,Giovanelli1991,Chengalur1995} and VIRGOHI 21 \citep{Davies2004,Minchin2005,Minchin2007,Haynes2007} that are extended sources with M$_\mathrm{HI} \sim 10^8-10^9$ \Msol, as well as many other clouds revealed in recent years, primarily by the Arecibo Legacy Fast Arecibo L-band Feed Array (ALFALFA) survey \citep{Kent2007} and the Arecibo Galaxy Environment Survey (AGES) \citep{Taylor2012}. Such dark features may be debris from tidal encounters between galaxies, or they may have their origin in the removal of gas from galaxies via ram pressure stripping by the intra-cluster medium (ICM).

With this in mind, we have begun the Widefield Arecibo Virgo Extragalactic Survey (WAVES). This is an extension of the Arecibo Galaxy Environment Survey (AGES) in the Virgo Cluster, with the specific intention of revealing the low column-density neutral hydrogen (\hi) content of the cluster. WAVES is currently covering the right ascension range $12^{\rm h}09^{\rm m}00^{\rm s}$ -- $12^{\rm h}49^{\rm m}00^{\rm s}$ and the declination range 09\degr00\arcmin00\arcsec -- 11\degr06\arcmin00\arcsec, filling in the gap between the AGES VC1 \citep{Taylor2012} and VC2 \citep{Taylor2013} fields to the same sensitivity and spectral resolution as AGES. 

The first quadrant of the survey, covering a 5 square degree area over RA $12^{\rm h}29^{\rm m}00^{\rm s}$ -- $12^{\rm h}49^{\rm m}00^{\rm s}$ and Dec. 09\degr00\arcmin00\arcsec -- 10\degr00\arcmin00\arcsec, was completed in March 2018. This takes in the area of the ALFALFA Virgo 7 complex, a dark \hi\ cloud complex discovered by the ALFALFA survey \citep[source 7 in][]{Kent2007} and identified as consisting of five clouds lying between 400 and 760 \kms\ (see also our Figure \ref{fig2}). VLA observations \citep{Kent2009} confirmed two of the clouds (C1 (7c) and C2 (7d)) but did not detect clouds C4 (7b) or C5 (7e) due to their distances from the pointing centre. Cloud C3 (7a) lay outside of the area mapped by the VLA. \citet{Sorgho2017} also covered this region with a combined WSRT and KAT-7 map; they detected cloud C1 but not cloud C4, which should have been detectable based on its single-dish \hi\ mass if it were a point source. The ALFALFA map of \citet[figure 2]{Kent2009} shows two of the five clouds in the complex (C1 and C4) to be connected, but the other three clouds are separated at their sensitivity level. They find a total mass for the complex of $5.1 \times 10^8$ \Msol\ at an assumed distance for the Virgo cluster of 16.7 Mpc, equivalent to a total flux of 7.8 \Jykms. \citet{Kent2009} found that a possible optical counterpart near the C2 cloud, VCC 1357, had a large separation from the VLA position and was thus unlikely to be associated with the complex.

The ALFALFA Virgo 7 complex is thus a unique structure in the Virgo cluster. Only one other \hi\ feature is known which is detached from its parent galaxy -- the cloud associated with VCC 1249 \citep{ArrigoniBattaia2012}. However, in that case the cloud is only 14 kpc projected distance from its parent, whereas the nearest plausible parent for the complex (NGC 4445) is separated by $\sim180$ kpc in projection from its center. The complex is also amongst the most massive of the streams known in Virgo and has a highly complex internal structure. No other feature in Virgo shows this combination of features. We describe our observations of the complex in Section \ref{Complex}.

This first quadrant also takes in the galaxy NGC 4522, a classic example of ram pressure stripping in the Virgo cluster \citep{Kenney1999}. This has a well-known \hi\ tail, observed by \citet{Kenney2004} and \citet{Chung2007} and simulated by \citet{Vollmer2006}. Like the ALFALFA Virgo 7 complex, the tail of NGC 4522 is an example of optically-dark \hi\ gas in Virgo although, unlike the complex, it remains connected to its parent galaxy and the mechanism by which it has been created is much clearer. We describe our observations of NGC 4522 in Section \ref{NGC4522}.

All velocities in this paper are radial velocities in the barycentric frame and the optical ({\it cz}) convention.
 
 \section{Observations}
 
Observations on this field were made between January 2017 and March 2018 using the Arecibo L-band Feed Array (ALFA). WAVES uses the same scan pattern and data reduction pipeline as AGES \citep[detailed in][]{Auld2006}. As around half of the data were taken prior to Hurricane Maria (20 September 2017), which caused changes to the gain of the Arecibo telescope, and half afterwards, separate cubes were initially made from each half of the dataset and their respective flux calibrations were compared using the fluxes of strong \hi\ sources within the cube. An adjustment of around 8 percent was then made to the post-Maria calibration and the post-Maria data were re-reduced with the new calibration before the final datacube was made.
 
 \subsection{Verification of the flux scale}
 
\begin{figure}
\plotone{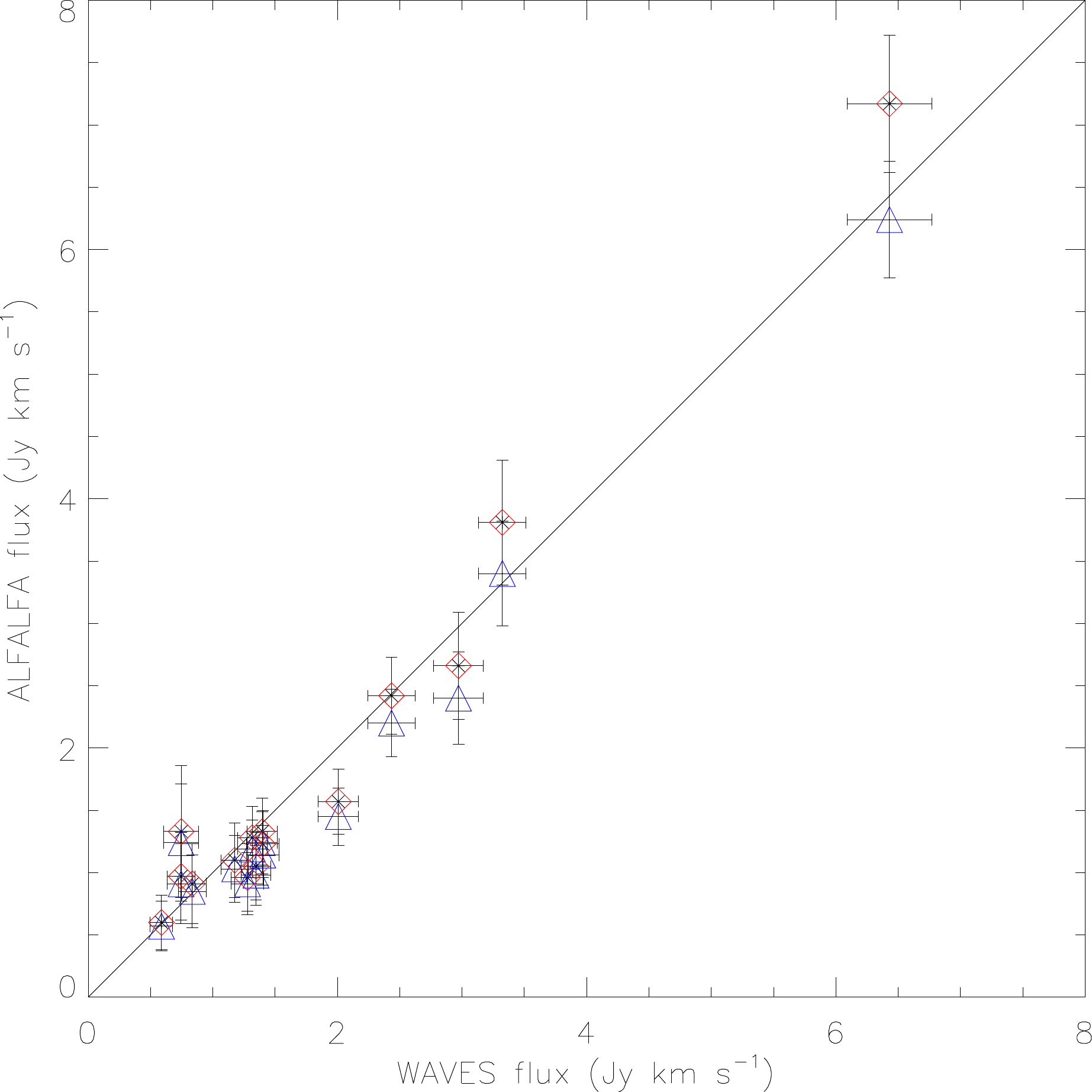}
\caption{Comparison of integrated \hi\ line fluxes (in \Jykms) of sources detected in both WAVES and ALFALFA. For ALFALFA, fluxes from \citet{Kent2008} are shown by blue triangles and fluxes from \citet{Haynes2018} are shown by red lozenges. Errors calculated according to the formulae of \citet{Koribalski2004}. To guide the eye, the solid line indicates a 1:1 relationship.\label{fig1}}
\end{figure}

We check our calibration by comparing our total \hi\ fluxes to those seen by ALFALFA. There are 33 sources in the ALFALFA catalogue of \citet{Kent2008} within the WAVES quadrant boundary. After removal of HVCs, the extended clouds from the complex, lower quality (`code 2') sources, and sources noted to have issues with their parameter fitting, we are left with 17 sources. Two of these lie too close to the edge of our cube to be well parameterized in WAVES, giving 15 good comparison sources. In Figure \ref{fig1} we compare our fluxes for these sources to the ALFALFA fluxes from \citet{Kent2008} and the rescaled ALFALFA fluxes from \citet{Haynes2018}, which are, on average, 8 percent higher. We find the weighted mean of the ratio of the WAVES flux to the \citet{Kent2008} flux is $1.07 \pm 0.04$ and for the \citet{Haynes2018} flux is $0.96 \pm 0.04$. Our flux scale thus appears to be slightly higher than the initial ALFALFA flux scale, but is consistent with their corrected flux scale.

 \section{The ALFALFA Virgo 7 complex}
 \label{Complex}
 
Around the velocity of the ALFALFA Virgo 7 complex, our observations reach an average noise level of 0.9 mJy beam$^{-1}$ at a Hanning-smoothed velocity resolution of 10 \kms, around 3 times deeper than the ALFALFA observations.

\begin{figure*}
\plotone{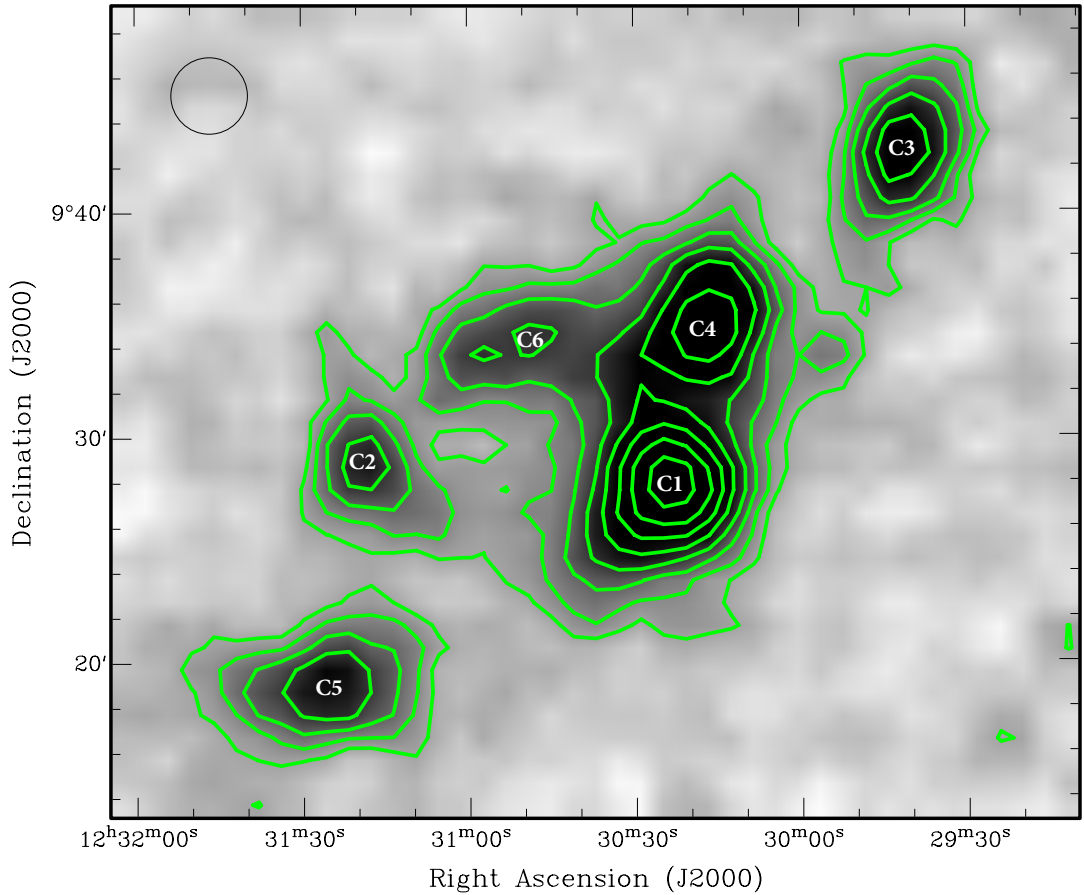}
\caption{Moment 0 map over 440 to 620 \kms with contours at 0.18, 0.30, 0.45, 0.65, 0.90, 1.20, 1.55, 1.95 \Jykms (equivalent to 5, 8, 12, 18, 24, 32, 42, 53 $\times 10^{19}$ cm$^{-2}$ for resolved structures), with clouds labeled. The velocity range is chosen to illustrate all 6 clouds at a good signal-to-noise and does not include all of the gas in the complex (see below). The Arecibo beam is shown by the black circle in the upper left corner. \label{fig2}}
\end{figure*}

Figure \ref{fig2} shows the moment 0 map of the complex over 440 to 620 \kms\, with the outer contour at $3\sigma$ (0.18 \Jykms, $5 \times 10^{18}$ cm$^{-2}$), the second contour at $5\sigma$ (0.30 \Jykms, $8 \times 10^{18}$ cm$^{-2}$), and other contours at increasing steps to bring out the peaks of the clouds. The five clouds identified in the complex by \citet{Kent2007,Kent2009} can be clearly seen, but we also see a sixth cloud (which we label C6 for consistency with \citealt{Kent2009}'s nomenclature) between C4 and C2. 

\begin{figure*}
\plotone{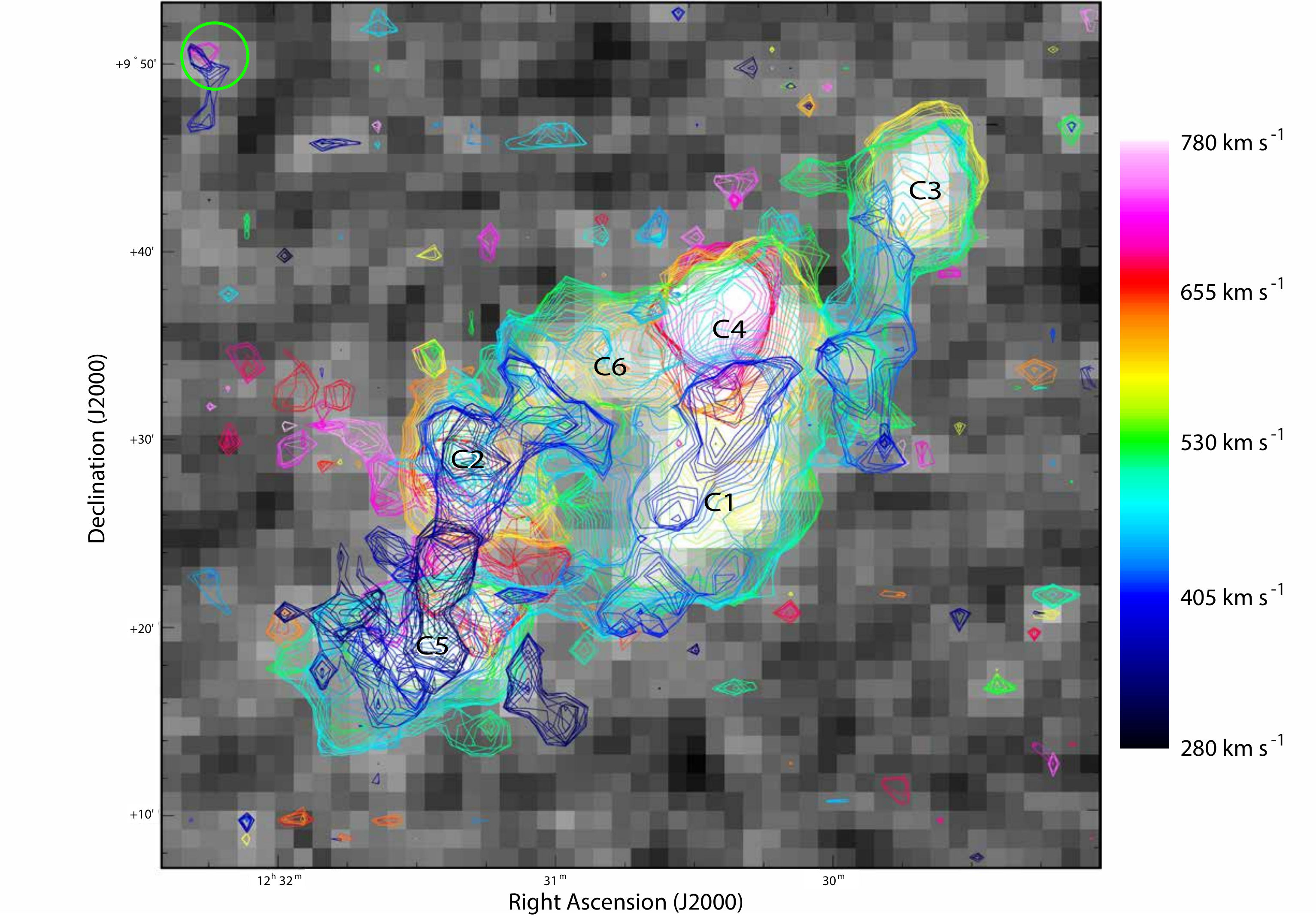}
\caption{Renzogram of the cube boxcar smoothed in velocity over a width of 11 channels to a velocity resolution of 55 \kms. Contours are at 1.5 mJy\,beam$^{-1}$ (equivalent to a column density of $2.2 \times 10^{18}$ cm$^{-2 }$ for resolved structures; note that this is much lower than the column densities in Figure \ref{fig2} due to the lower velocity range covered by each contour). Colors indicate velocity, using a `Rainbow' color map from 280 \kms\ (violet) to 780 \kms\ (red). A renzogram can be thought of as over-plotted color-coded channel maps, allowing the connections between structures at different velocities to be seen much more easily than with a traditional channel map and at a much higher resolution than would be possible with a large grid of traditional channel maps. Greyscale shows the moment 0 map over the velocity range 250--800 \kms. The green circle in the top left indicates the beam size. \label{fig3}}
\end{figure*}

Figure \ref{fig3} shows channel maps averaged over 55 \kms\ and color-coded by velocity, over the range 280 to 780 \kms. It is apparent that, unlike in the map of \citet{Kent2009}, at the WAVES sensitivity level the whole complex is joined up by \hi\ bridges. We also see gas down to 280 \kms, 120 \kms\ below the lower limit of the gas seen by \citet{Kent2009}; this is shown by violet and purple contours in Figure \ref{fig2}. The new cloud C6 is visible in the cyan and green contours, and partly in the yellow contours. 

\begin{figure*}
\plottwo{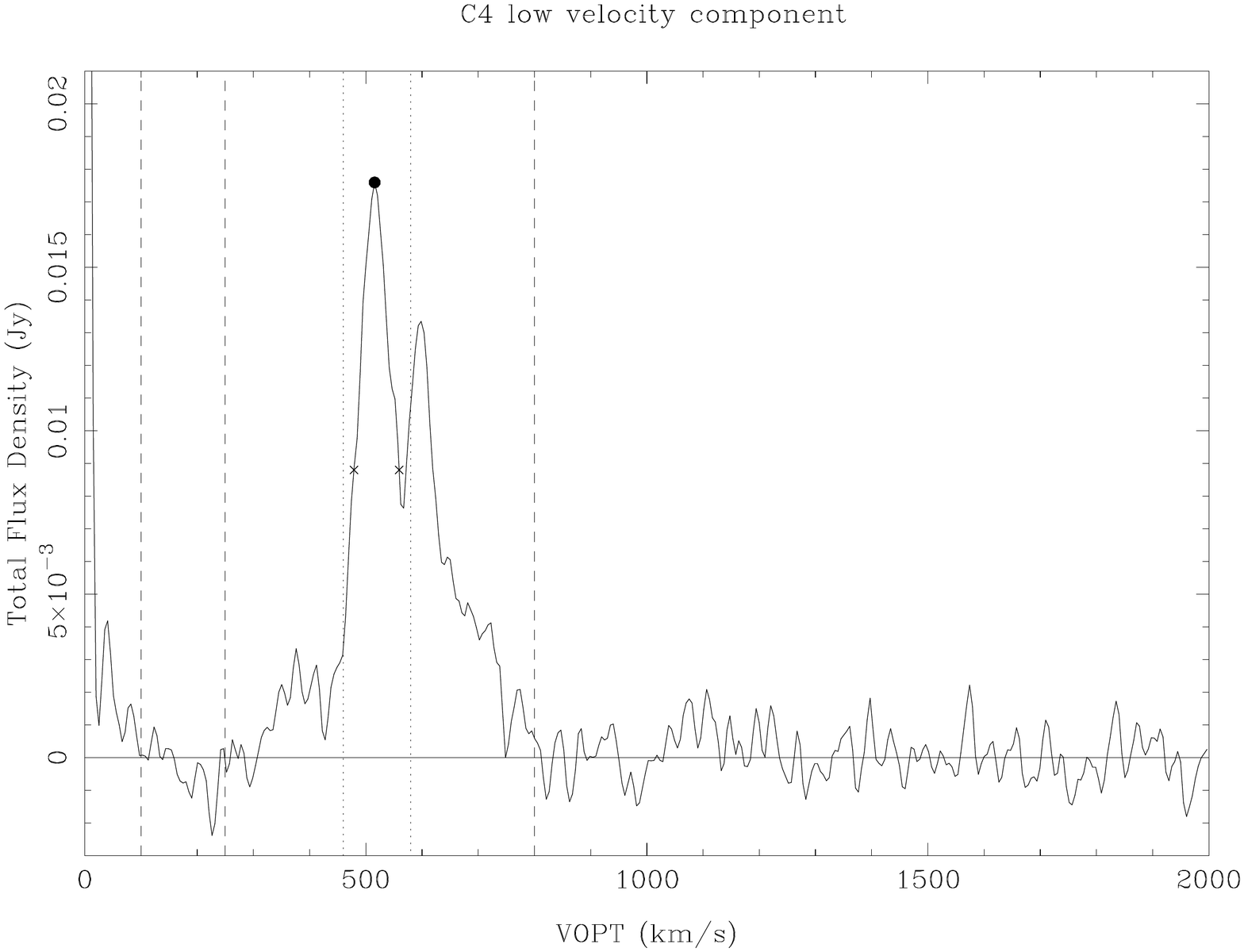}{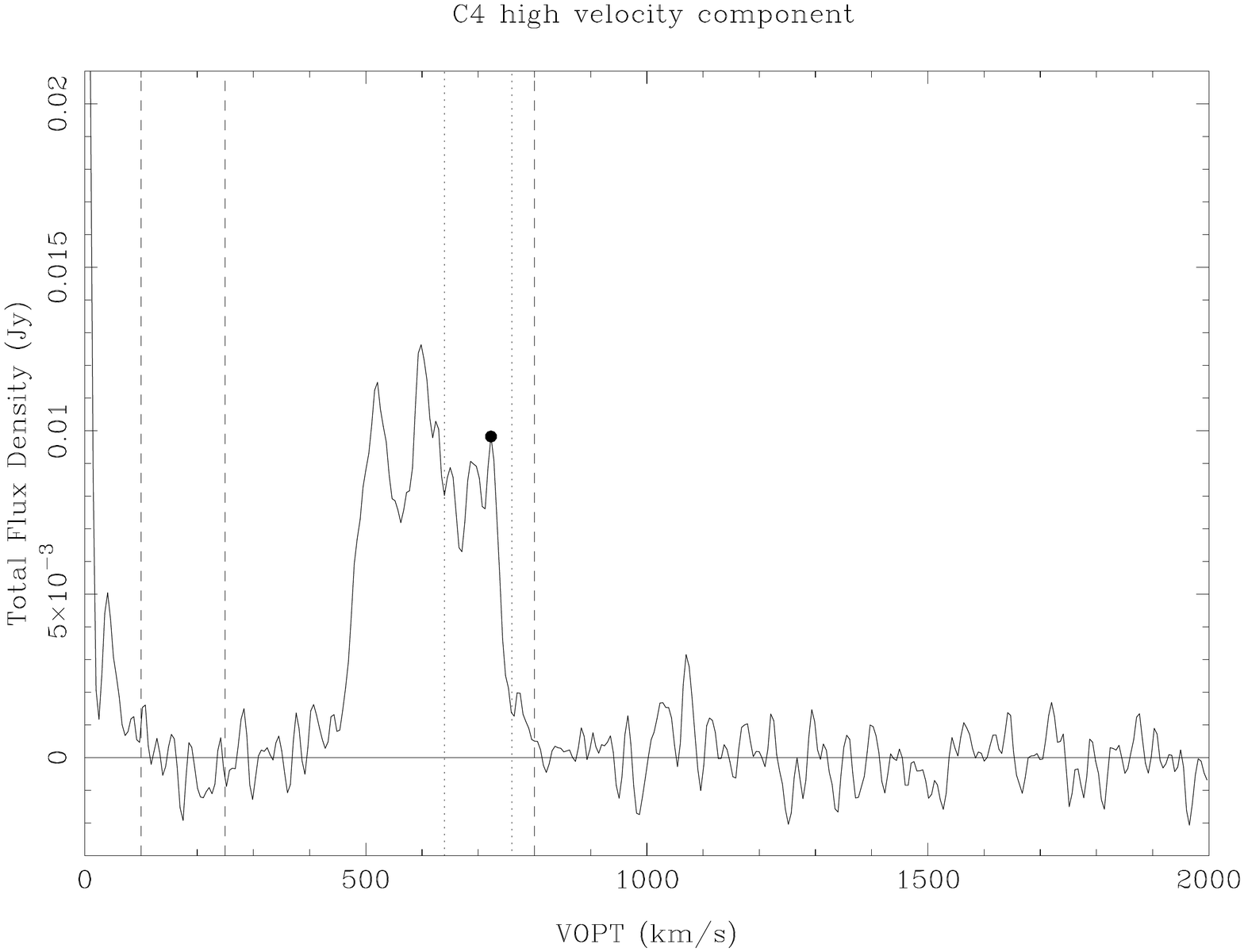}
\caption{Spectra of the low velocity (left) and high velocity (right) components of C4, summed and beam-corrected over a $5\arcmin \times 5\arcmin$ region. Dotted lines mark out the velocity range used for the position fit. A first order baseline has been subtracted from both spectra; the region over which this was fitted is indicated by the extent of the spectra dashed lines indicate the ranges excluded from the baseline fit (below 100 \kms\ to exclude Galactic hydrogen and 250--800 \kms\ to exclude the complex itself).\label{fig4}}
\end{figure*}

Cloud C4 can be clearly seen to consist of two overlapping components, with a shift in the centre of around 2.5\arcmin between the cyan and green contours and the orange and red contours. The higher velocity component has a position (fitted over 640 to 760 \kms) of $12^{\rm h}30^{\rm m}24.0^{\rm s}$, 09\degr36\arcmin50\arcsec while the lower velocity component has a position (fitted over 460 to 580 \kms) of $12^{\rm h}30^{\rm m}17.7^{\rm s}$, 09\degr34\arcmin40\arcsec. Spectra of the two components (summed over a $5\arcmin \times 5\arcmin$ box) are shown in Figure \ref{fig4}.

\begin{figure}
 \plotone{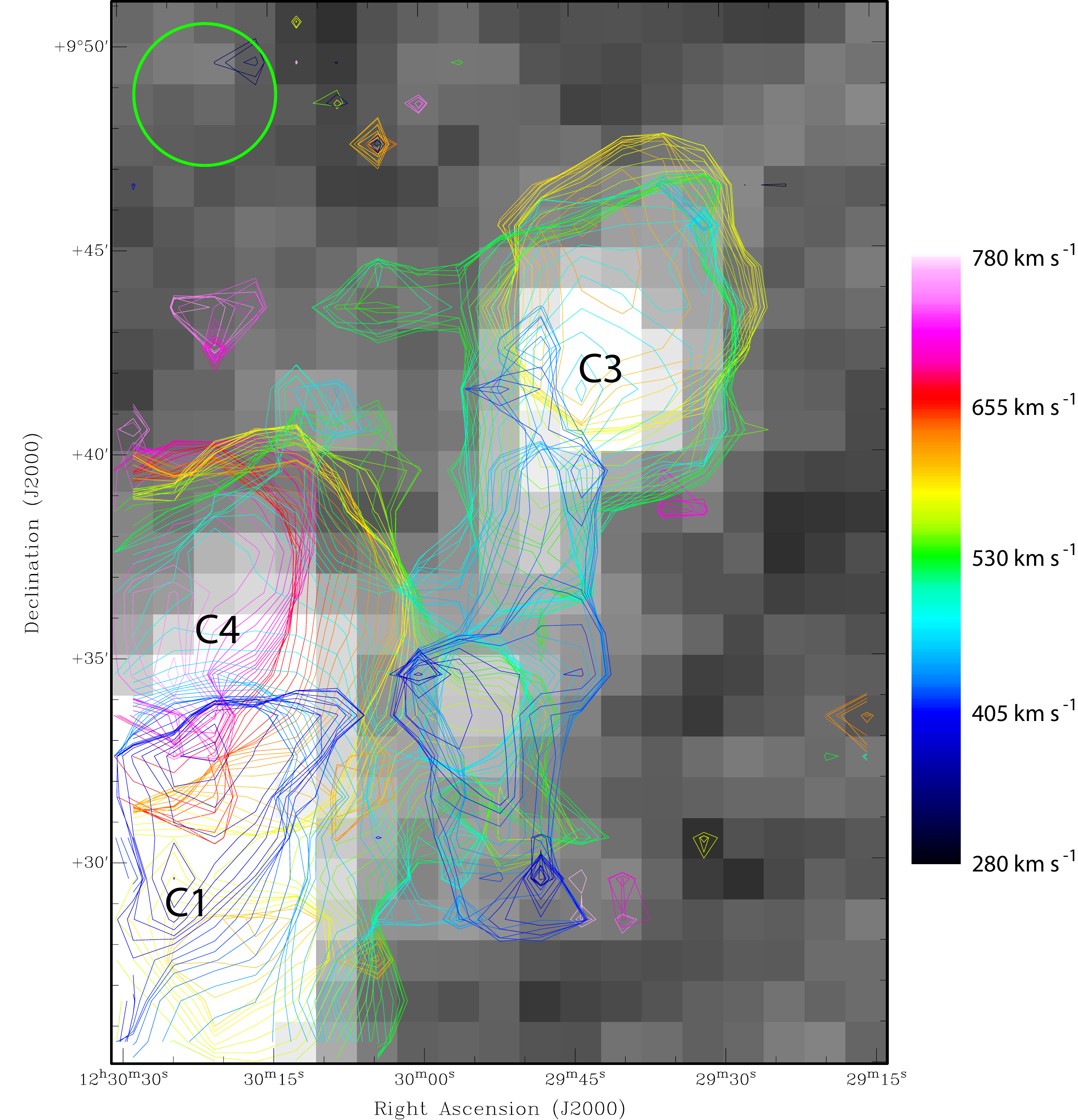}
 \caption{Zoomed region of the renzogram in Figure \ref{fig3} showing the bridge between C3 and the main (C1-C4) complex. Greyscale shows a moment 0 map over 400 -- 520 \kms. The green circle in the top left indicates the beam size.\label{fig5}}
 \end{figure}

\citet{Kent2009} describe cloud C3 as `unresolved with the ALFA beam'. However, we see the cloud as being clearly extended along an axis pointing towards C1, with the centre shifting further along this axis (away from C1) at higher velocities. This can be seen in Figure \ref{fig5}. A bridge between cloud C3 and the main complex can be seen in the blue and cyan contours and in the greyscale on this figure, which covers the same velocity range. The bridge connects to an extension seen at low SNR to the west side of C4 in the \citet{Kent2009} map, which is now seen to be real (conversely, the extension to the northeast side of C4, seen at a very similar flux level in the \citealt{Kent2009} map, is not seen in our data). However, in velocity space the connection looks to be to C1 (which would be consistent with the alignment of the axis of cloud C3) rather than to C4.
 
\begin{figure}
\plotone{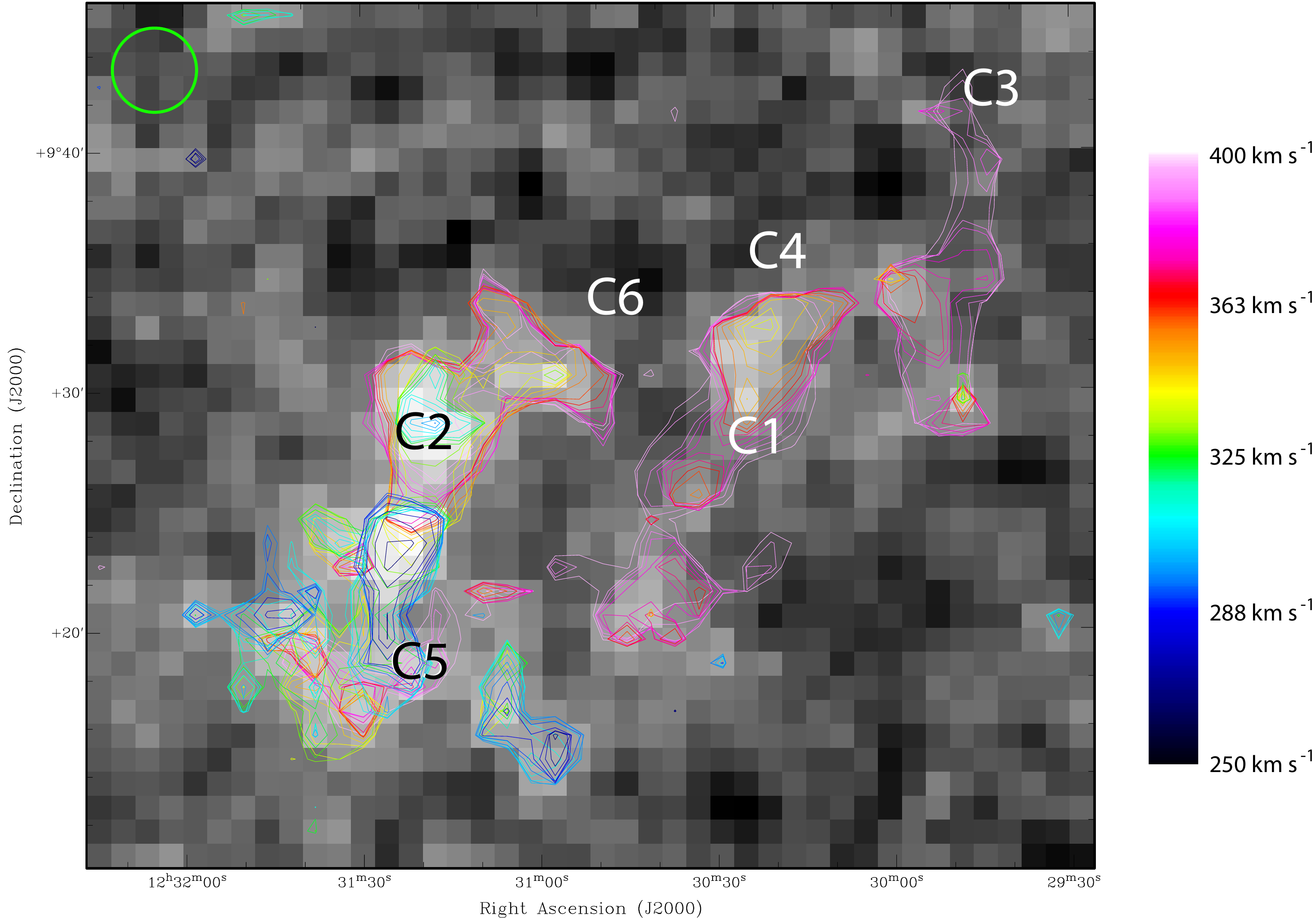}
\caption{Renzogram (rainbow colormap; contour levels as in Figure \ref{fig3}) and moment 0 map (greyscale) of low velocity gas over 250--400 \kms. The green circle in the top left indicates the beam size.\label{fig6}}
\end{figure}

Clouds C2 and C5 contain lower velocity gas than that seen by \citet{Kent2009}, which bridges between the two clouds; this is marginally visible (at around $2\sigma$) in the \citet{Kent2009} spectrum of C2 but is clearly detected here. This is shown in Figure \ref{fig6}. It can be seen that there is detected gas at the positions of C5 and C2, stretching about half way to the position of C6 and with the peak in the 280--340 \kms\ map about half way between C5 and C2. There is also gas in the main complex, with a peak between the positions of C1 and C4, stretching to the root of the bridge to C3. The gas in this velocity range does not contribute greatly to the total \hi\ content of the complex, with F$_\mathrm{HI} = 0.34$ \Jykms\ (after sidelobe correction, see below), for an \hi\ mass of $2.3 \times 10^7$ \Msol, i.e. less than 2 percent of the total \hi\ mass of the complex. Nevertheless, it provides important information on the connections between the clouds.

The `bay' between C2 and C1, south of the new C6 cloud, is not completely filled in at any of the velocity slices, even though almost all of its area is covered at some velocity or other (see Figures \ref{fig2} and \ref{fig3}). We fit a position for the C6 cloud (using a $5\arcmin \times 5\arcmin$ box) of $12^{\rm h}30^{\rm m}46.8^{\rm s}$, 09\degr34\arcmin04\arcsec. The flux in the single-pixel spectrum at this position is $0.78 \pm 0.12$ \Jykms, with a peak of only 6 mJy, making it less than $3\sigma$ in the \citet{Kent2007,Kent2009} data, while the beam-corrected flux within the $5\arcmin \times 5\arcmin$ box is $0.97 \pm 0.13$ \Jykms. The total flux measured off the moment 0 map within a $7\arcmin \times 8\arcmin$ box enclosing C6 is $1.62 \pm 0.12$ \Jykms.

\subsection{Total \HIit\ content of the complex}

\begin{figure*}
\plottwo{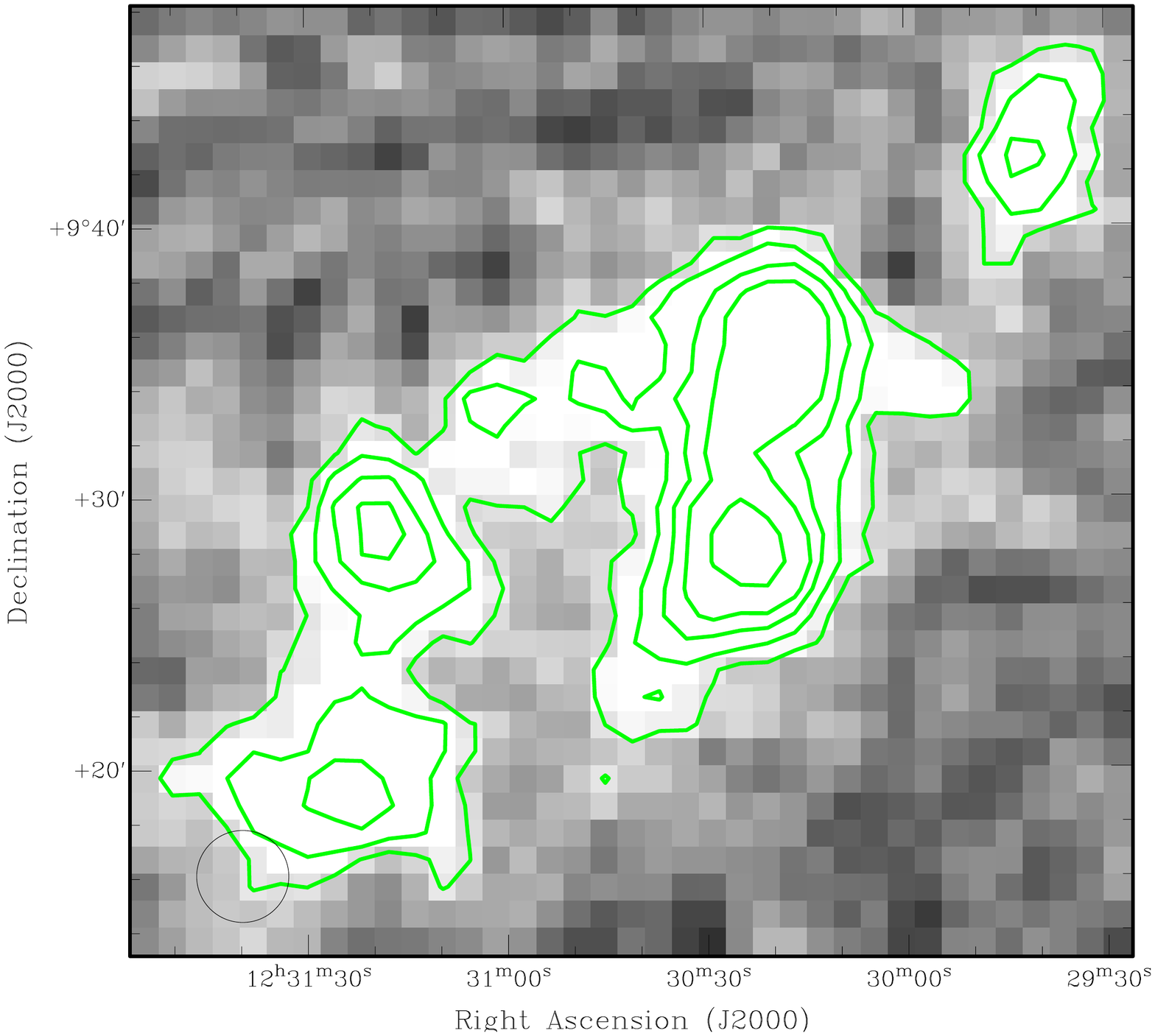}{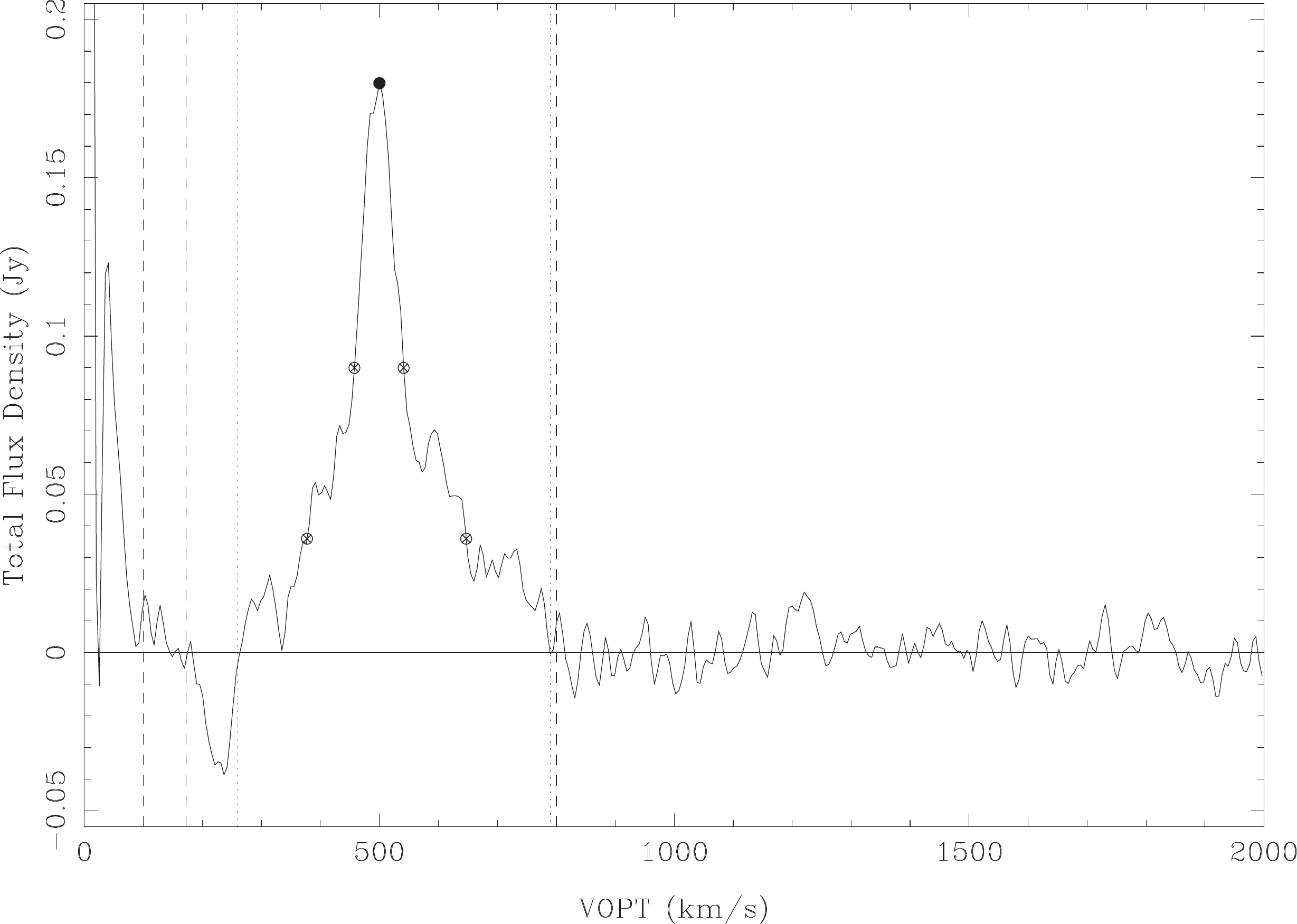}
\caption{Left: Moment 0 map over 260 to 790 \kms\ of the complex over the $37\arcmin \times 35\arcmin$ region centered on $12^{\rm h}30^{\rm m}41^{\rm s}$, 09\degr30\arcmin38\arcsec used for total flux measurements, with contours showing 3$\sigma$, 5$\sigma$, 7.5$\sigma$, 10$\sigma$, 15$\sigma$ (over the full velocity width, so narrow features such as the bridge between C1/C4 and C3 are washed out by noise from other channels). Right: Spectrum over the same area showing the measurement range (dotted line). A first order baseline has been subtracted from the spectrum; the region over which this was fitted is indicated by the extent of the spectrum, dashed lines indicate the ranges excluded from the baseline fit (below 100 \kms\ to exclude Galactic hydrogen and 180--800 \kms\ to exclude the complex itself and the nearby negative feature); the negative feature at 180 to 260 km/s is an artefact from high velocity Galactic hydrogen in the eastern part of the cube.\label{fig7}}
\end{figure*}

We measure the \hi\ flux of the complex in two ways: by measuring the total flux in a moment 0 map summed spectrally across the 260--790 \kms\ velocity range in a $37\arcmin \times 35\arcmin$ region centered on $12^{\rm h}30^{\rm m}41^{\rm s}$, 09\degr30\arcmin38\arcsec (fully enclosing points above $3\sigma$ on the moment 0 map) and by measuring the total flux over the same velocity range in a spectrum summed spatially across the same region (these are shown in Figure \ref{fig7}). These are corrected using a circular Gaussian beam of HPBW 3.4\arcmin. The moment 0 map gives a total flux of $26.6 \pm 1.3$ \Jykms\ and the spectrum gives a total flux of $28.0 \pm 1.7$ \Jykms. While these are not significantly different, they are different enough that using a polygonal map to better fit the shape of the complex is unlikely to bring a major improvement in signal to noise while potentially missing some diffuse gas. Combining the measurements from the two methods gives a value of $27.3 \pm 2.2$ \Jykms. 

This total flux measurement needs to be adjusted for the sidelobes, which are not included in the  Gaussian primary beam correction, and which contain contain significant flux scattered out of the main beam. As the first minimum occurs at $\sim 3.7\arcmin$ from the beam center and the peak of the first sidelobe is at $\sim 5.6\arcmin$, for a measurement box size larger than $5\arcmin \times 5\arcmin$ at least part of the first sidelobe will fall within the box and will thus lead to an overestimate of the flux if this is not corrected for (a sidelobe correction is not necessary for normal point-source measurements in AGES as these are made within a $5\arcmin \times 5\arcmin$ box, falling entirely within the $\sim 3.7\arcmin$ radius of the first minimum and thus not including any sidelobe flux). A box of size $15\arcmin \times 15\arcmin$ or greater should fully enclose the first sidelobe, thus our map (which is substantially larger than this) will contain both mainbeam flux and flux from the sidelobes, both of which will contribute to our measured flux. \citet{Heiles2001} found the ratio of the first sidelobe and main beam efficiencies to be $\eta_\mathrm{FS}/\eta_\mathrm{MB} = 0.33$ for the LBW feed, while \citet{Heiles2004} found values for the ALFA beams of $\eta_\mathrm{FS}/\eta_\mathrm{MB} = 0.16$ for the central beam and varying from 0.30 to 0.38 for the outer beams. The average across all ALFA beams is $\eta_\mathrm{FS}/\eta_\mathrm{MB} = 0.31$ and the median is 0.33. We adopt a value of $\eta_\mathrm{FS}/\eta_\mathrm{MB} = 0.33$ and correct our measured flux by a factor of $1 + \eta_\mathrm{FS}/\eta_\mathrm{MB}$, giving $20.5 \pm 1.7$ \Jykms. This assumes no significant contribution from other sidelobes: as $\langle\eta_\mathrm{MB} + \eta_\mathrm{FS}\rangle = 0.75$, sidelobes beyond the first contain around 25 percent of the total flux, but the majority of this will be scattered well outside our map; we therefore follow \citet{Peek2011} in not attempting a correction for the more distant sidelobes.

We therefore measure a total \hi\ mass of $1.3 \pm 0.1 \times 10^9$ \Msol, assuming the same distance of 16.7 Mpc used by \citet{Kent2009} This is considerably higher than their total \hi\ mass measurement of $5.1 \times 10^8$ \Msol. However, their total flux is simply the sum of the flux found in each cloud, to which we have added substantial extended emission and emission from smaller clouds that could explain the discrepancy. To check this, we compare the measurements from single-point spectra at the locations \citet{Kent2009} identify as the cloud centers. For consistency, the errors for \citet{Kent2009} were re-calculated using the formulae of \citet{Koribalski2004} and the quoted S/N.

We find that our measurements of the individual clouds are consistent with \citet{Kent2007,Kent2009}. There are no significant differences in fluxes on individual clouds between the WAVES single-point spectra and the \citet{Kent2009} measurements, and summed across the five clouds identified by \citet{Kent2009} we see less emission than they do, although not significantly ($7.20 \pm 0.30$ vs $7.83 \pm 0.92$ \Jykms; errors following \citet{Koribalski2004}). Thus the extra flux seen in our map would appear to be from extended emission and newly identified clouds not included in the total by \citet{Kent2009}. The only significant difference between our measurements and \citet{Kent2009} is on the central velocity and velocity width of cloud C3. This inconsistency can be traced to a narrow secondary peak near 600 \kms\ that is above 50 percent of the flux of the primary peak in the \citet{Kent2009} spectrum, thus increasing their 50 percent velocity width and shifting their velocity center, but which falls below 50 percent of the peak of the primary in the corresponding single-point WAVES spectrum (the peak in question can be seen in our spatially-integrated spectrum of C3 in Figure \ref{fig8}, where it is again above 50 percent).

\subsection{\HIit\ parameters of individual clouds} 
 
\begin{figure*}
\plottwo{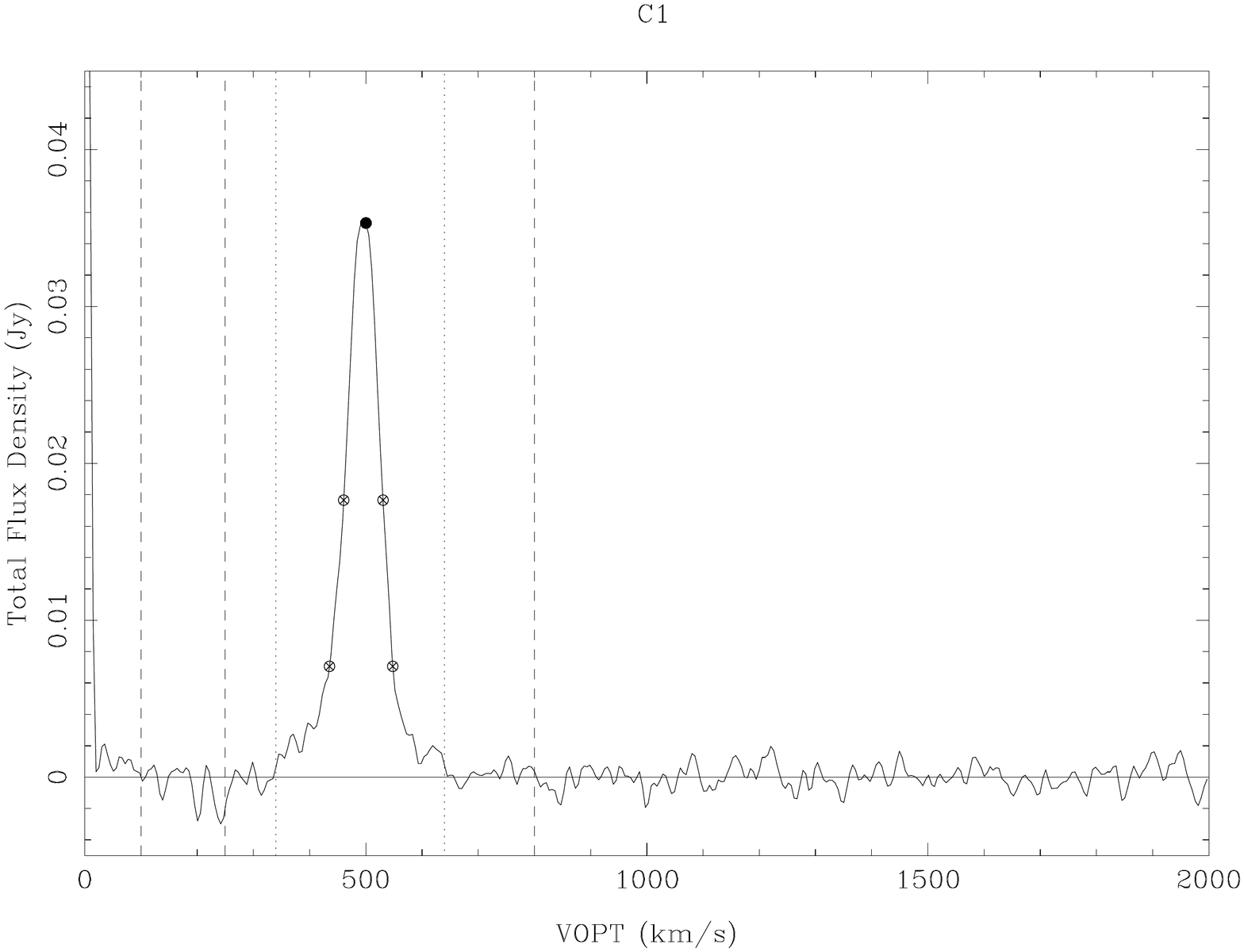}{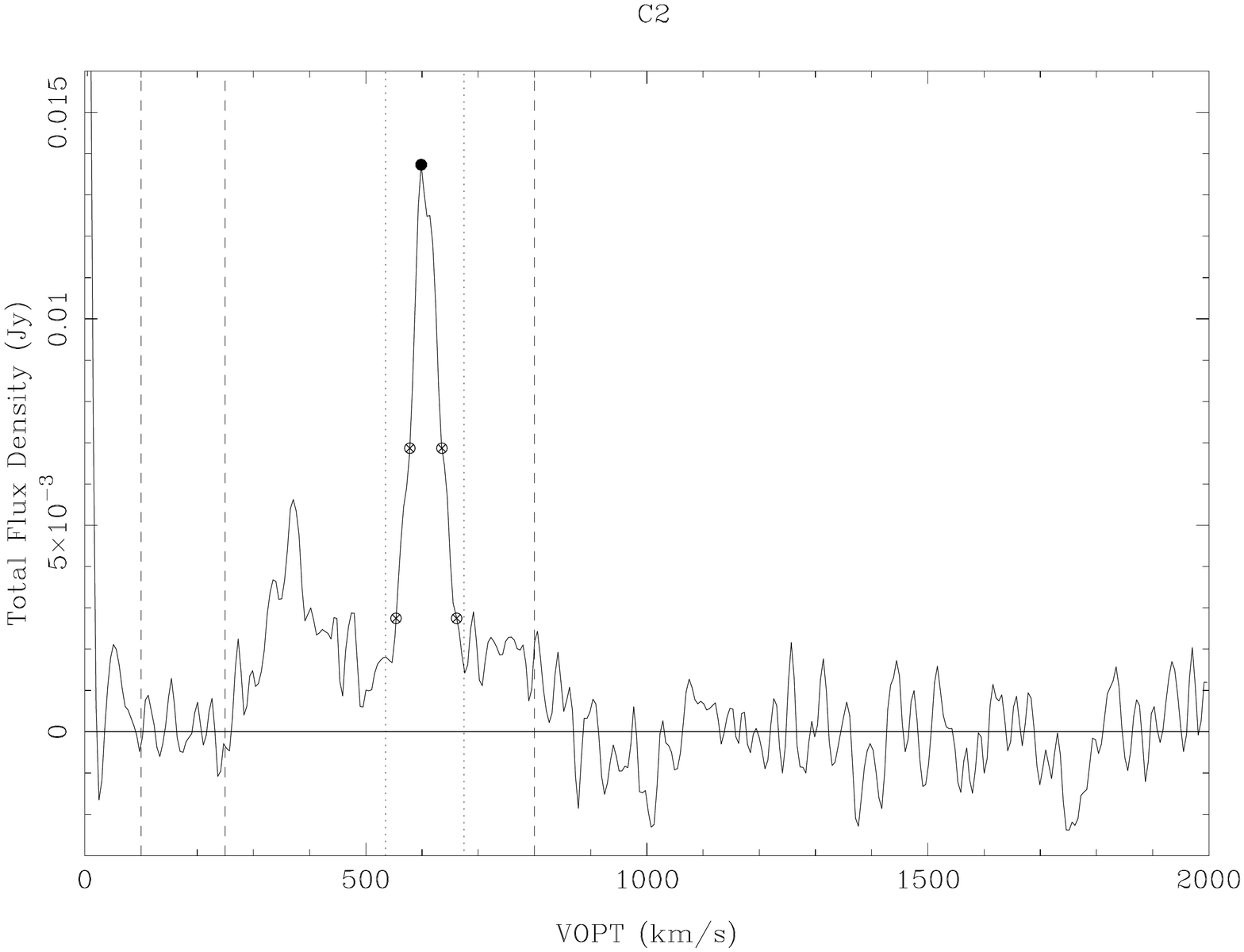}
\plottwo{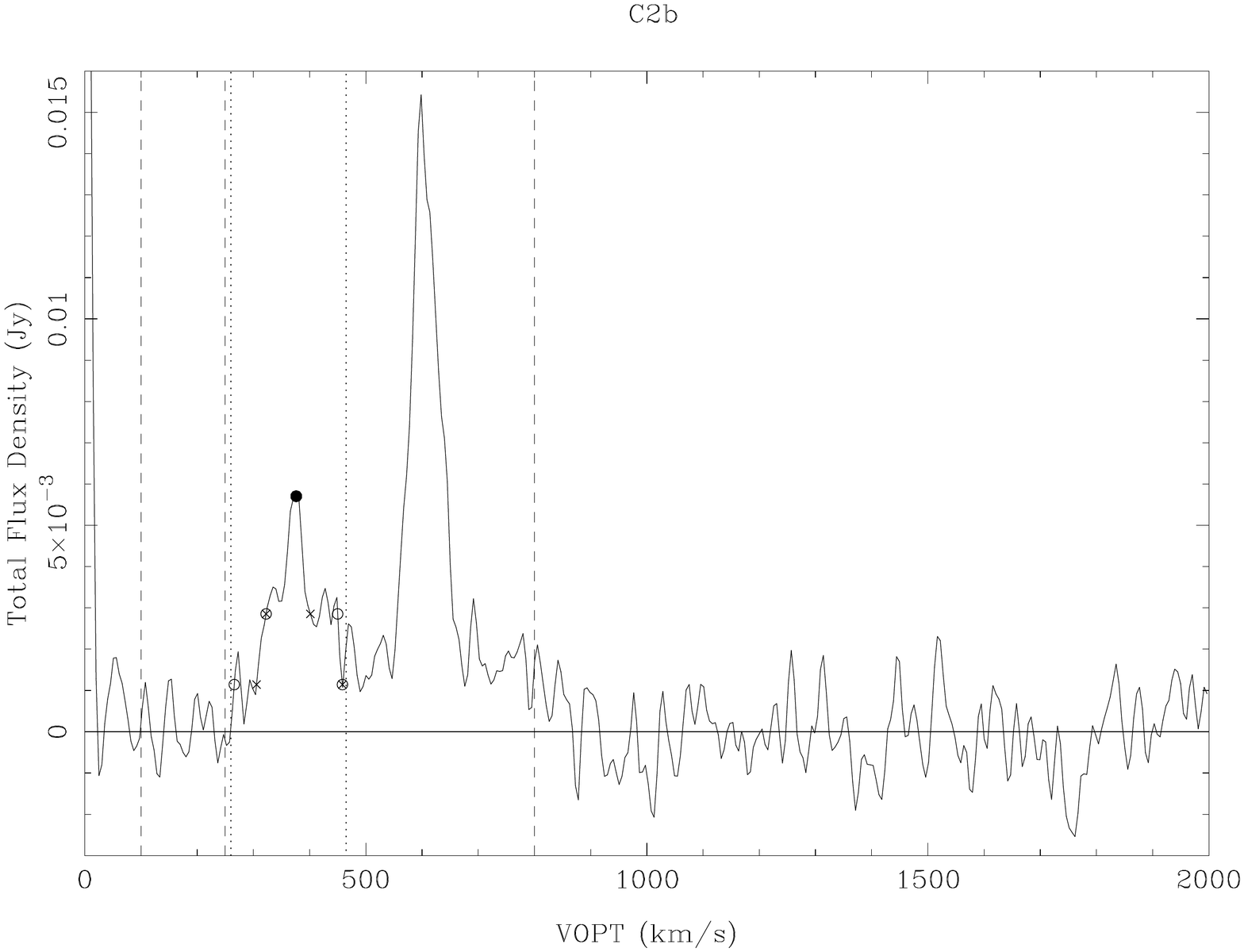}{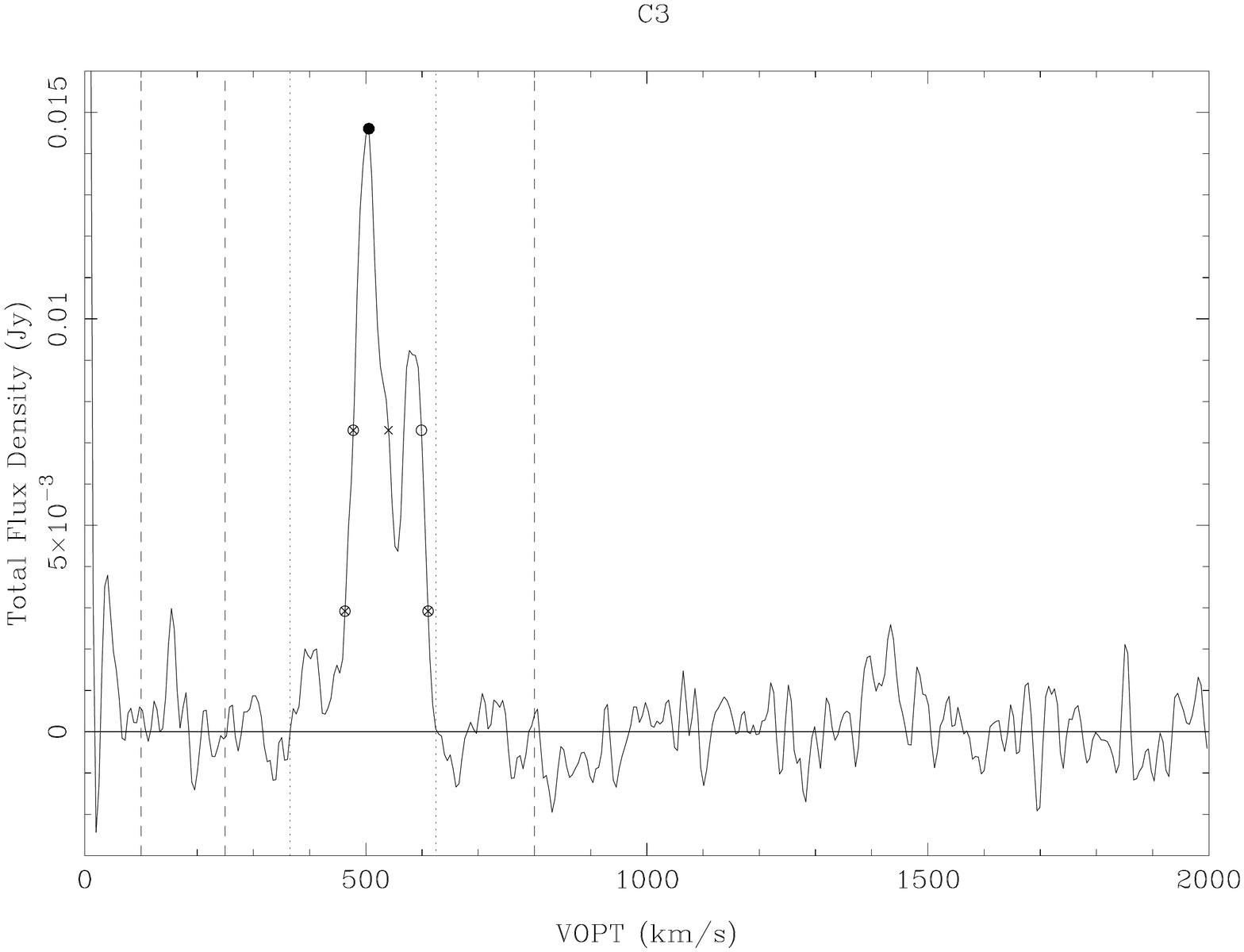}
\plottwo{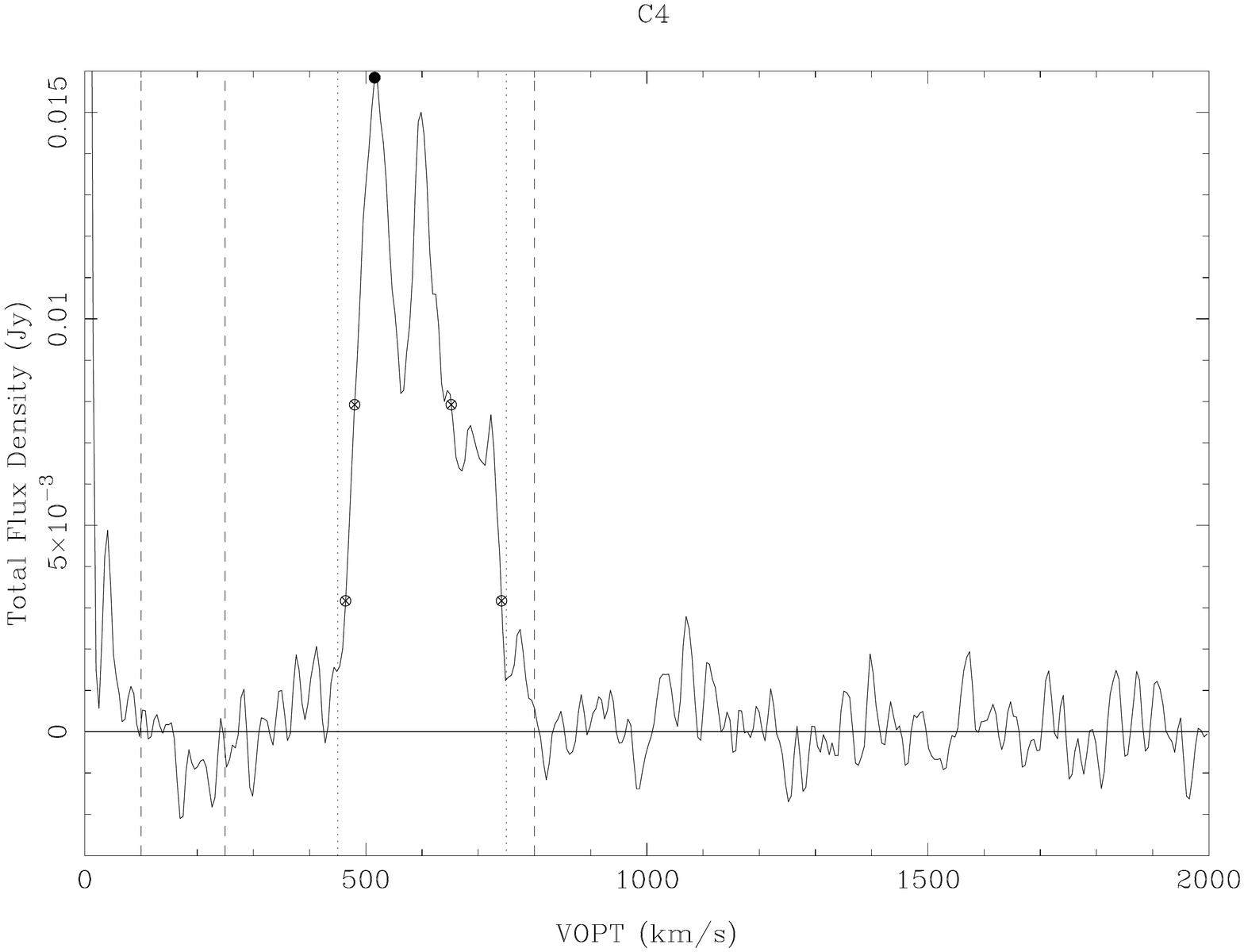}{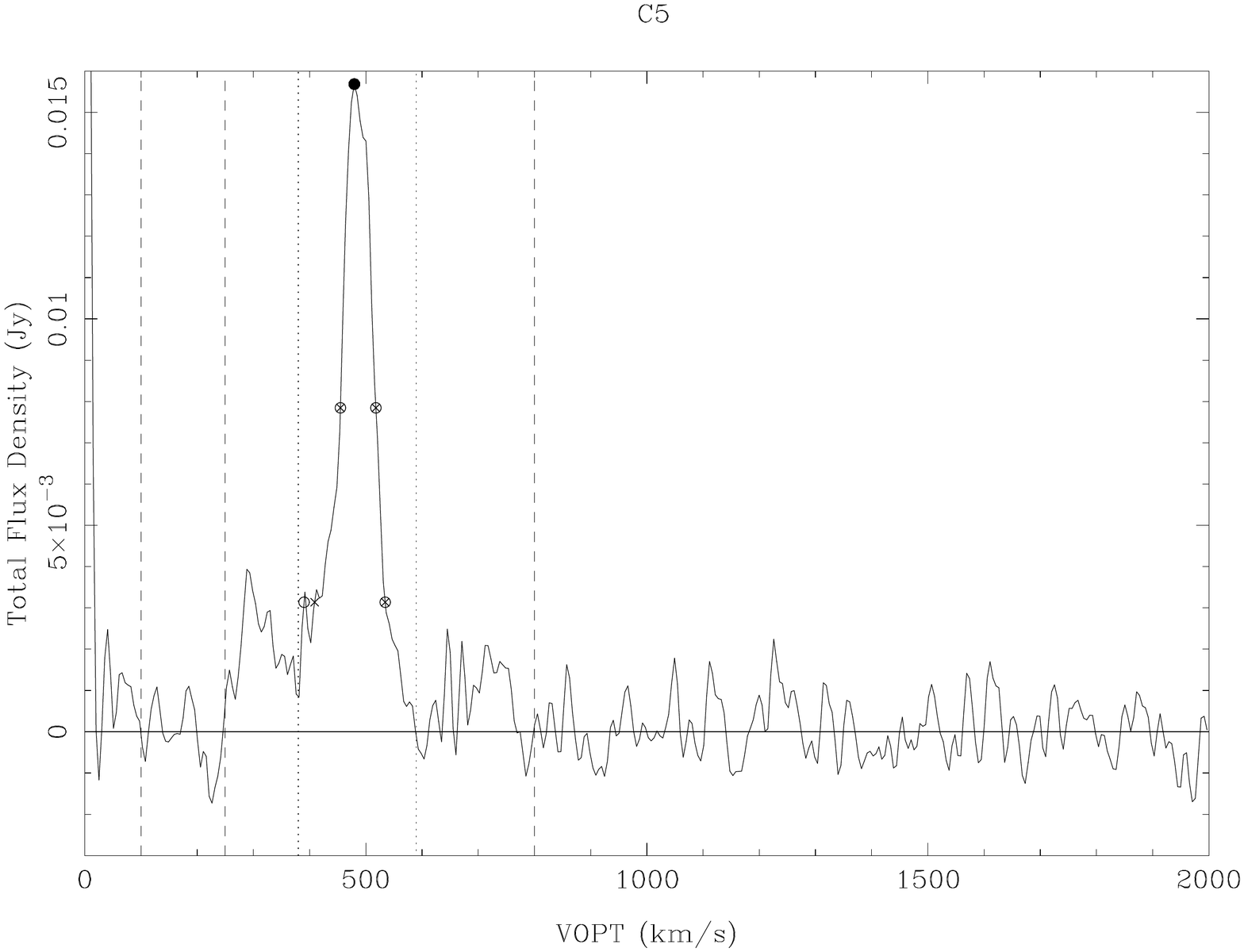}
\plottwo{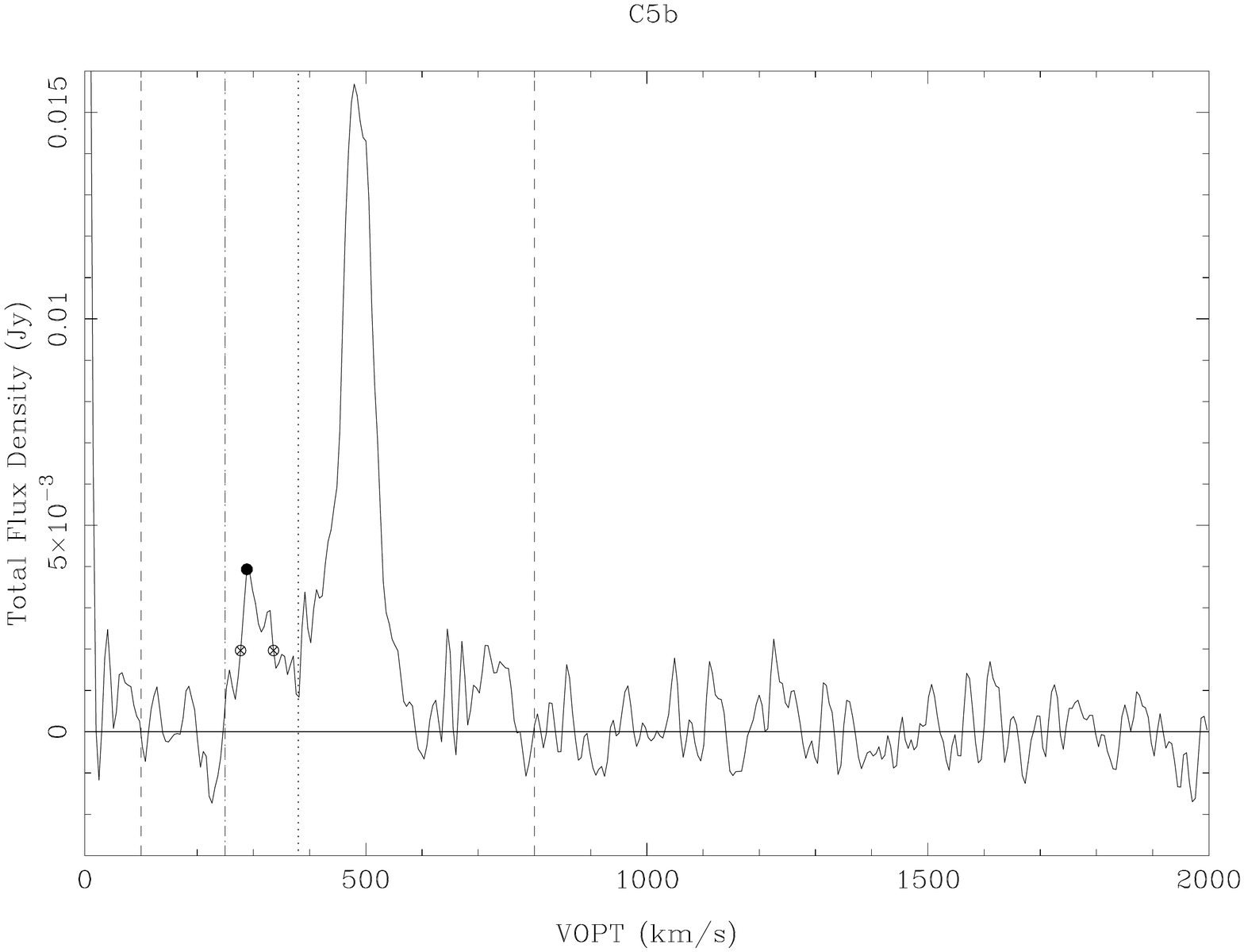}{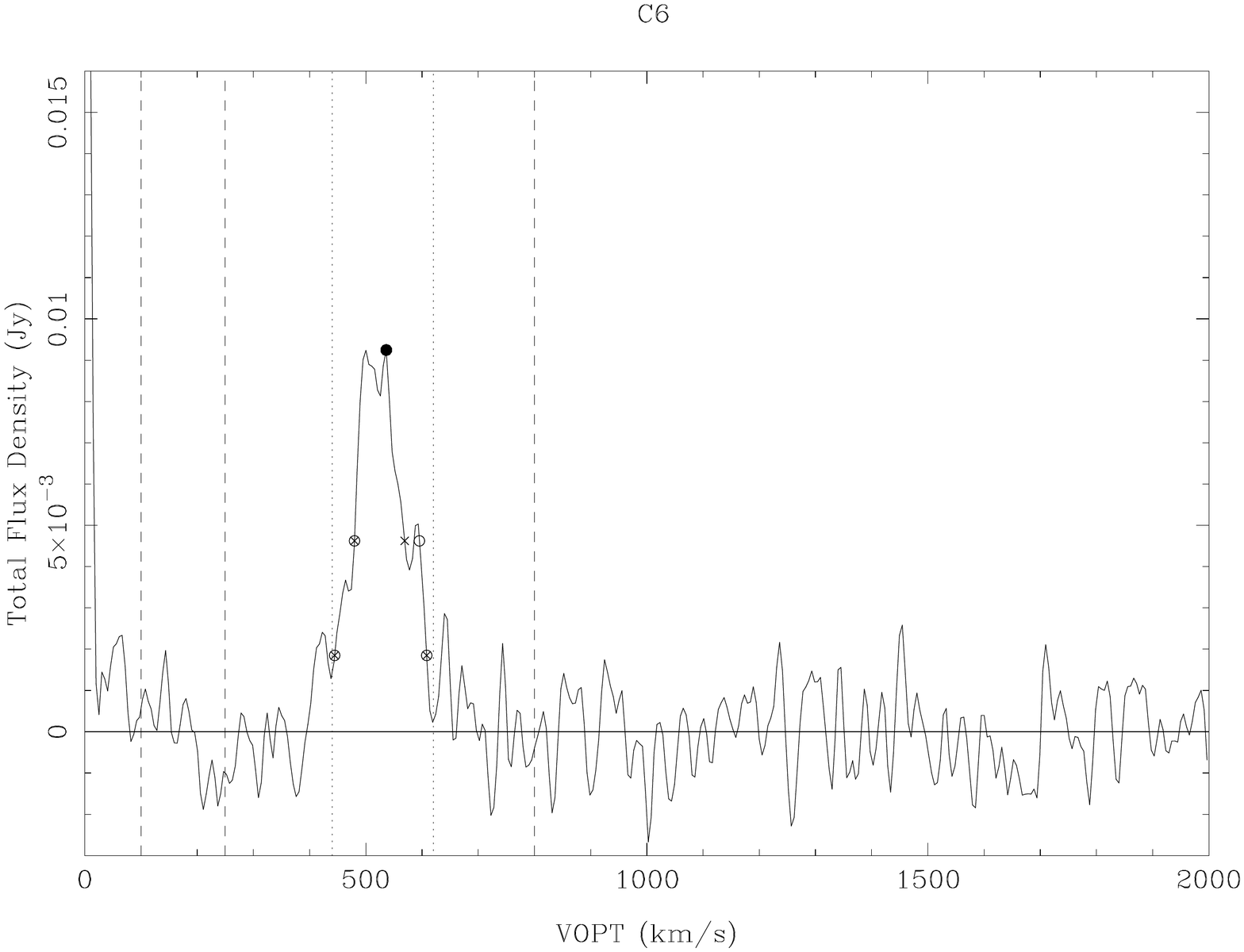}
\caption{Spectra of the clouds used for the measurements in Table 2. Left to right, top row: C1, C2; second row: C2b, C3; third row: C4, C5; bottom row: C5b, C6. For all spectra, dotted lines indicate the velocity range used to derive the parameters given in Table \ref{table1}. The region used for the baseline fit is indicated by the extent of the spectra; dashed lines indicate velocity ranges excluded from the fit due to possible contamination from the complex or Galactic gas.\label{fig8}}
\end{figure*}

Measuring over a $5\arcmin \times 5\arcmin$ box allows us to fit positions for the clouds in addition to measuring their parameters (Table \ref{table1}). In addition to the 5 clouds found by \citet{Kent2007,Kent2009}, we also fit for cloud C6 and the new low-velocity components of clouds C2 and C5, here labeled C2b and C5b, which are separated kinematically from the clouds found by \citet{Kent2009}. Although cloud C4 has two distinct center positions these are not clearly separated kinematically; thus we only make a single measurement of the flux, velocity width, etc. for that cloud. The spectra associated with these measurements are given in Figure \ref{fig8}.

\begin{deluxetable*}{llllllllll}
\tablecaption{Measurements of the clouds in a $5\arcmin \times 5\arcmin$ box. Errors calculated using the formulae of \citet{Koribalski2004}.\label{table1}}
\tablehead{
ID&RA&Decl.&V$_{50}$&W$_{50}$&W$_{20}$&F$_\mathrm{HI}$&M$_\mathrm{HI}$&S$_\mathrm{peak}$&Peak\\
&(J2000)&(J2000)&(\kms)&(\kms)&(\kms)&(\Jykms)&($10^8$ \Msol)&(mJy)&SNR}
\startdata
C1&12:30:24.6&09:27:59&$496 \pm 2$&$70 \pm 5$&$113 \pm 7$&$3.06 \pm 0.22$&$2.01 \pm 0.14$&35&67\\
C2&12:31:19.2&09:28:40&$607 \pm 3$&$	57 \pm 7$&$108 \pm 10$&$0.94 \pm 0.10	$&$0.62 \pm 0.06$&14&18\\
C2b&12:31:19.9&09:28:59&$386 \pm 7$&$127 \pm 14$&$192 \pm 21$&$0.57 \pm 0.09$&$0.37 \pm 0.06$&6&8.0\\
C3&12:29:42.2&09:42:39&$538 \pm 2$&$	121 \pm 4$&$148 \pm 7$&$1.40 \pm 0.12$&$0.91 \pm 0.08$&15&23\\
C4&12:30:19.9&09:35:47&$566 \pm 4$&$	172 \pm 8$&$278 \pm 12$&$2.73 \pm 0.16$&$1.79 \pm 0.11$&16&29\\
C5&12:31:25.8&09:18:49&$486 \pm 4$&$	63 \pm 8$&$144 \pm 11$&$1.27 \pm 0.11$&$0.83 \pm 0.07$&16&27\\
C5b&12:31:28.7&09:18:29&$307 \pm 8$&$57 \pm 15$&$107 \pm 23$&$0.28 \pm 0.07$&$0.18 \pm 0.05$&4&6.7\\
C6&12:30:47.3&09:34:04&$538 \pm 5$&$	115 \pm 11$&$164 \pm 16$&$0.97 \pm 0.13$&$0.64 \pm 0.09$&9&11
\enddata
\end{deluxetable*}

The gas clouds identified in Table \ref{table1} contribute a total flux of $11.2 \pm 0.4$ \Jykms, compared with our measurement of $20.5 \pm 1.7$ \Jykms\ for the complex as a whole. Diffuse gas in between the compact clouds would thus appear to be responsible for at least $45 \pm 4$ percent of the total gas in the complex, with the strong possibility that more exists below our detection threshold. Unlike in the single-point spectra discussed above, the fluxes in these spatially-extended spectra are higher than those found by \citet{Kent2009} for the same clouds, indicating the likely presence of extended flux around these sources.

 \section{NGC 4522}
 \label{NGC4522}

\begin{figure*}
\plottwo{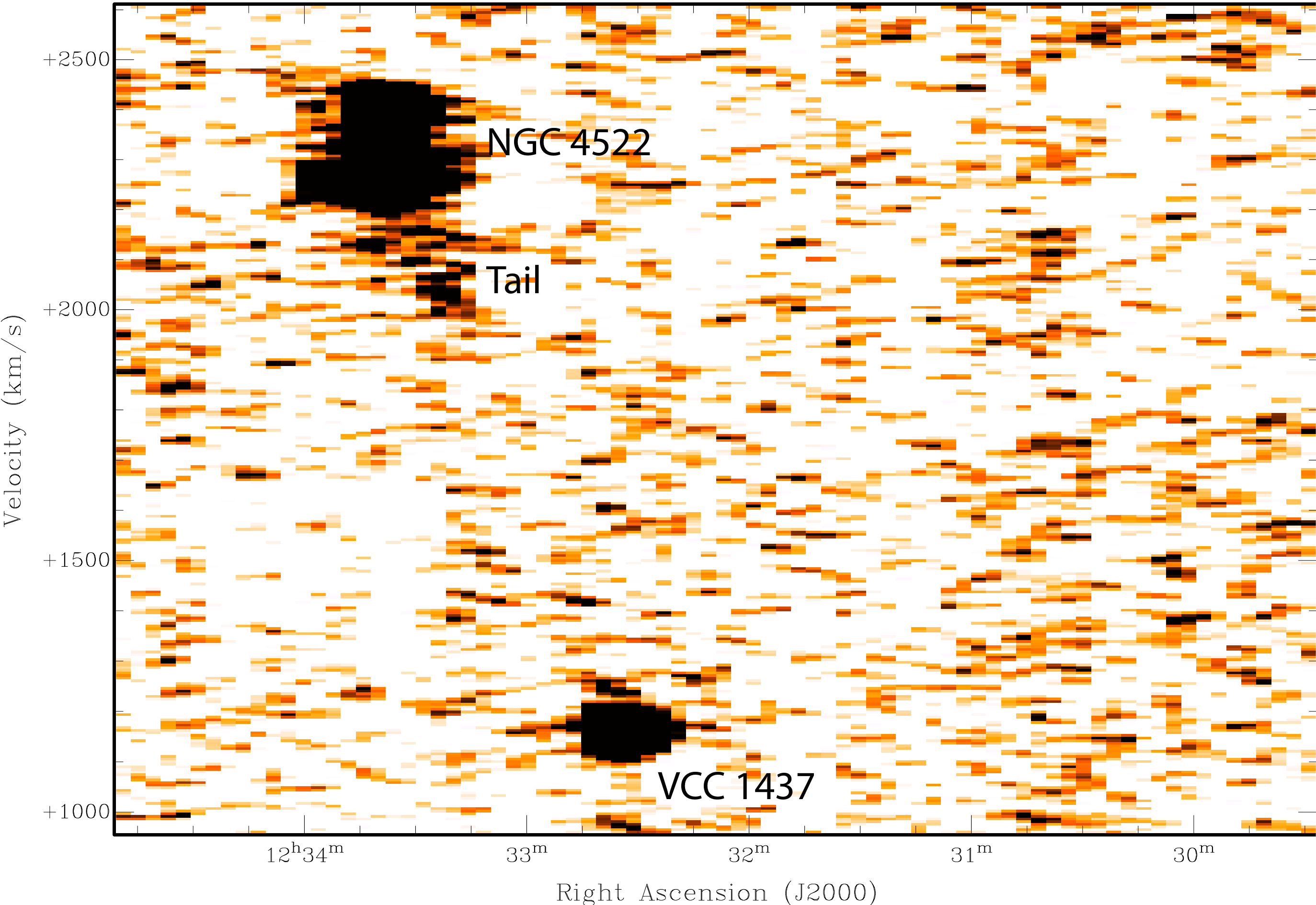}{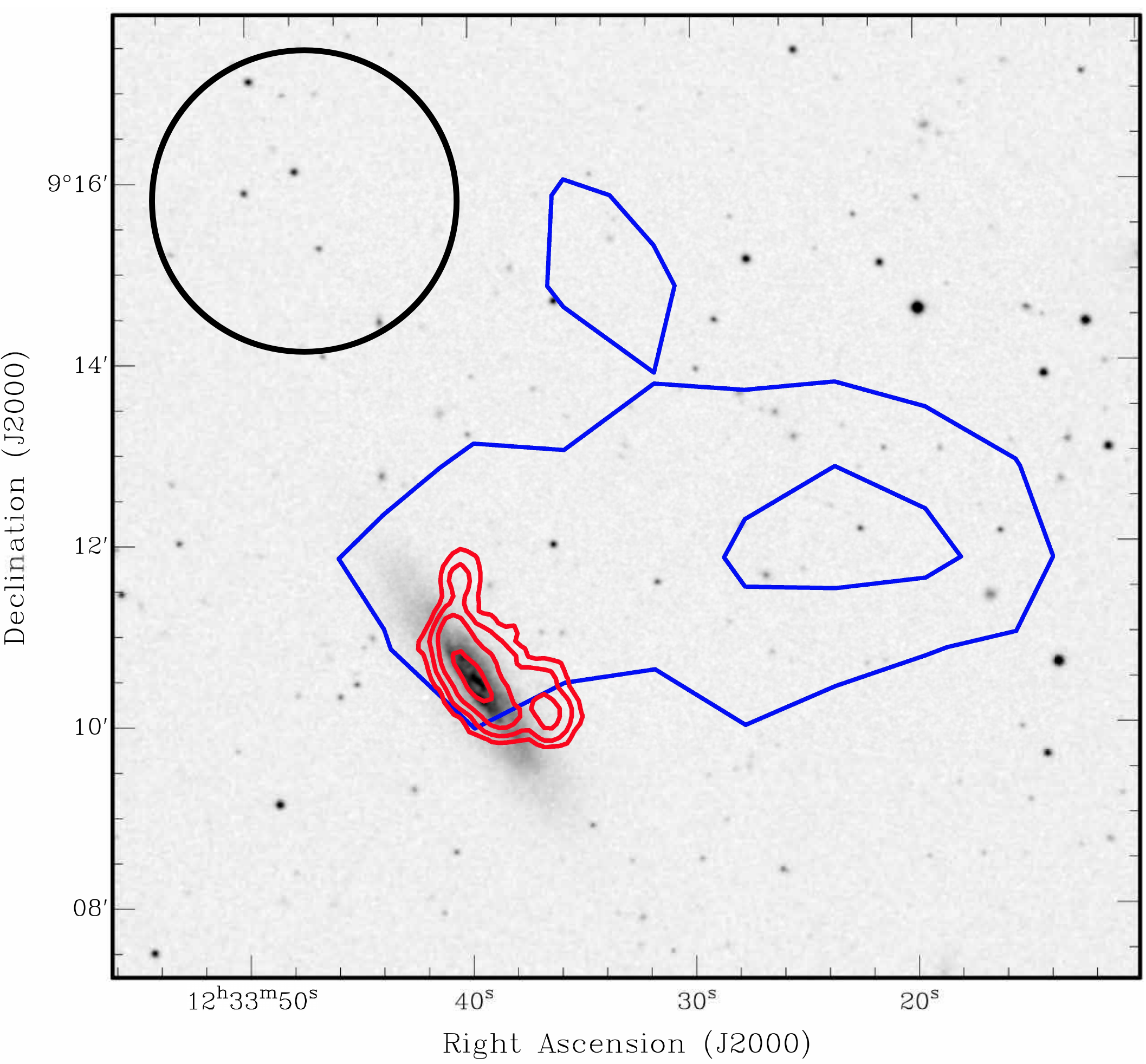}
\caption{Left: Position-velocity map showing velocity as a function of RA, covering the declination range 09\degr11\arcmin to 09\degr13\arcmin (the range over which the tail is visible in the data cube), showing the tail to low velocities on NGC 4522 and the galaxy VCC 1437 . Right: Gray scale shows the DSS image of the galaxy with blue contours showing the WAVES moment 0 map covering 2000 to 2200 \kms\ in velocity (i.e. only velocities below those of the main body of the galaxy) at $5 \times 10^{18}$ cm$^{-2}$ and $1 \times 10^{19}$ cm$^{-2}$, illustrating the eastwards extension of the tail from NGC 4522, and red contours showing the VLA moment 0 map from VIVA \citep{Chung2007}, covering around 2170 to 2470 \kms, at $2 \times 10^{20}$, $5 \times 10^{20}$, $1 \times 10^{21}$ and $2 \times 10^{21}$ cm$^{-2}$. The Arecibo beam is shown to the top left.\label{fig9}}
\end{figure*}

NGC 4522 is a classic example (see Figure \ref{fig9}) of a galaxy that is undergoing ram-pressure stripping, giving rise to an optically-dark \hi\ tail of gas removed from the galaxy. VLA \hi\ observations by \citet{Kenney2004} and \citet{Chung2007} show a tail of around 40 arcsec in length beyond the plane of the galaxy; \citet{Kenney2004} further derive an \hi\ deficiency of 0.6, indicating that around three quarters of the original gas content has been lost from this galaxy. \citet{Vollmer2006} successfully simulate the extra-planar \hi\ and the \hi\ deficiency by assuming the galaxy is traveling at a high speed ($\sim 3500$ \kms) relative to the ICM, either because it is not bound to the cluster or because the ICM is itself moving with respect to the cluster due to the infall of the M49 group. In their simulations, the galaxy has an initial gas mass of $1.3 \times 10^9$ \Msol, of which $9 \times 10^8$ \Msol\ is lost. 

\begin{figure}
\plotone{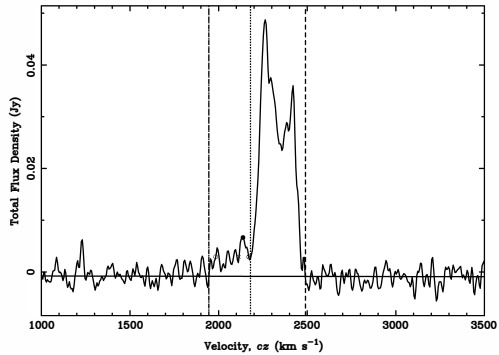}
\caption{Spectrum of NGC 4522 summed over a $13\arcmin \times 7\arcmin$ box centered on $12^{\rm h}33^{\rm m}30^{\rm s}$, 09\degr12\arcmin00\arcsec and primary beam corrected. A first order (linear) baseline has been fitted and is shown; the extent of the spectrum indicates the region over which this fit was made. Dashed lines at 1945 and 2490 \kms indicate the area masked from the baseline fit and dotted lines at 1945 and 2180 \kms indicate the area used for the measurement of the tail.\label{fig10}}
\end{figure}

To this, our new observations add the discovery of a high velocity (relative to the galaxy) extension of the gas tail of NGC 4522. This stretches about 200 \kms\ below the velocity of the \hi\ seen in the galaxy, to a recessional velocity of 2000 \kms, and has a slight inclination seen in a velocity-RA map (Figure \ref{fig10}), shifting around 25 seconds in RA (6\arcmin = 30 kpc at a distance of 16.7 Mpc) west along its length to 13:33:15. The \citet{Vollmer2006} simulations reproduced the blueshift of the \hi\ velocities in the tail relative to the galaxy seen in the VLA observations of \citet{Kenney2004}, but not the velocity width seen there of 150 \kms\ (full width at zero intensity). Our addition of an extra 200 \kms\ to this is thus well out of the range of gas dynamics reproduced by the simulations. The direction of the tail we observe is consistent with the angle of $-15\deg$ (north of west) assumed by \citet{Vollmer2006} based on HST observations.

The extension of the tail can be seen in the spectrum of NGC 4522 as a faint wing on the low velocity side of the main peak (Figure \ref{fig10}). From measurements on the spectrum, summed over a $13\arcmin \times 7\arcmin$ box centered on $12^{\rm h}33^{\rm m}30^{\rm s}$, 09\degr12\arcmin00\arcsec, the wing has an \hi\ flux of $0.9 \pm 0.2$ \Jykms\ within a velocity range from the minimum between the galaxy and the tail at 2180 \kms\ to when it falls to zero intensity at 1945 \kms. This gives an \hi\ mass (at 16.7 Mpc) of $5.9 \pm 1.3 \times 10^7$ \Msol. By comparison we measure (on the same spectrum) a flux of $7.9 \pm 0.5$ \Jykms\ for the galaxy between the minimum at 2180 \kms\ and when it falls to zero intensity at 2490 \kms\, equivalent to $5.2 \times 10^8$ \Msol\ (due to the size of the box, this will include a small contribution from the sidelobes). Measurement assuming a point source at the position of the \hi\ peak of the galaxy gives a flux of $6.4 \pm 0.3$ \Jykms, equivalent to $4.2 \times 10^8$ \Msol. These are both consistent with the $7 \pm \sim 1$ \Jykms\ measured in a single Arecibo pointing by \citet{Helou1984}, and bracket the $7.17 \pm 0.10$ \Jykms\ measured by ALFALFA \citep{Haynes2018}.

Allowing for beam smearing, the WAVES tail extends around 5\arcmin from the galaxy to a column density of $\sim 10^{19}$ cm$^{-2}$. This is $\sim 7.5$ times longer than the tail visible on the VIVA maps from the VLA \citep{Chung2007}, over which distance the column-density falls by a factor of $\sim 20$. The tail density therefore decreases approximately as $N_{\mathrm HI} \propto d^{-1.5}$. This is consistent with geometrical dilution being the dominant process in making the tail invisible, as inferred from comparison with results of \citet{Roediger2007} and from computations of test particle models as in \citet{Koeppen2018}.

\section{Discussion -- possible origins of the ALFALFA Virgo 7 complex}
 
\begin{figure*}
\plotone{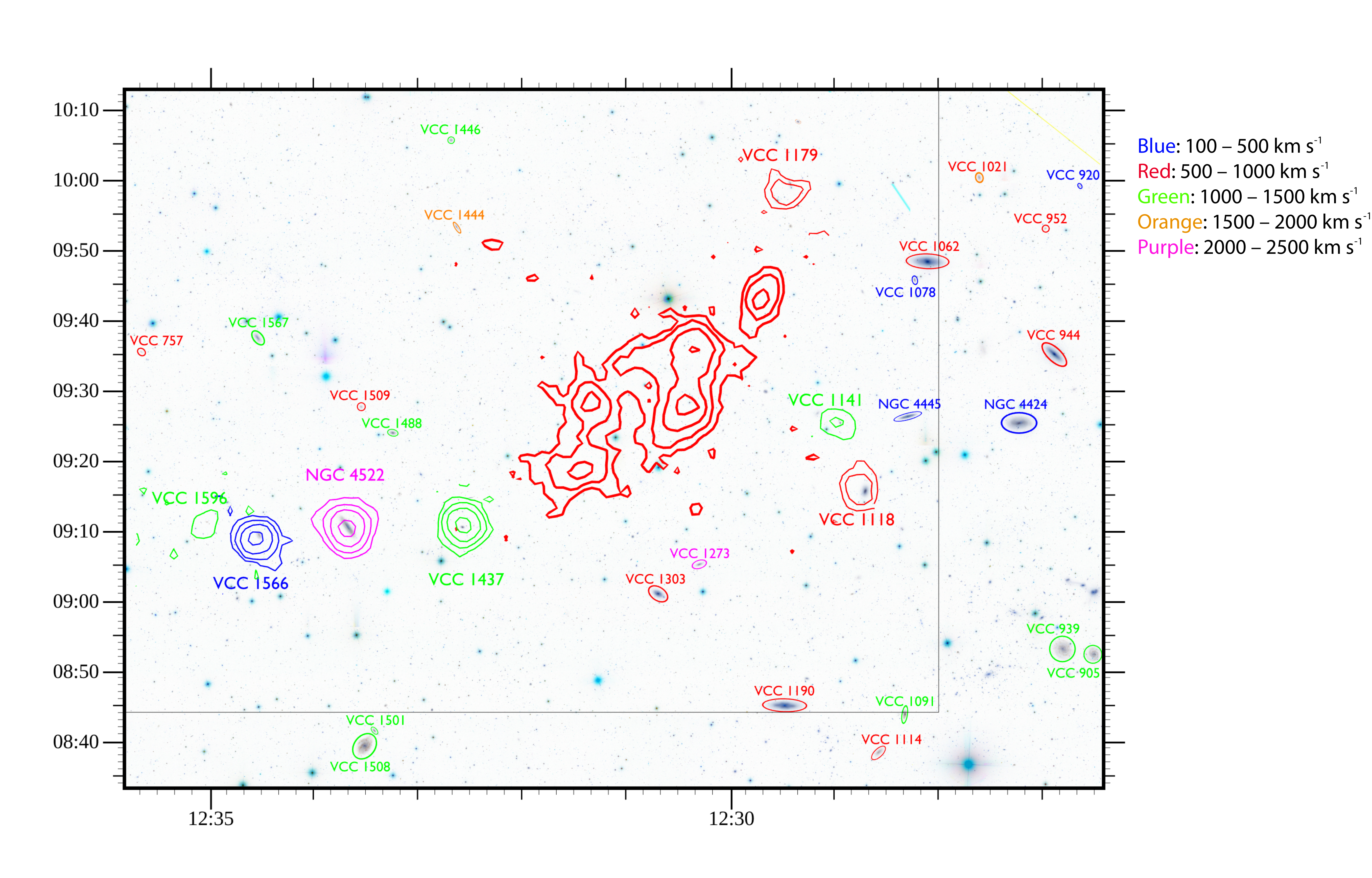}
\caption{Significant galaxies near the ALFALFA Virgo 7 complex. \hi\ contours from WAVES are shown for those galaxies detected (contour levels vary between sources due to their different fluxes and velocity widths and are set to ensure the individual sources are visible); others (undetected or outside of the current WAVES coverage) are shown by ellipses. Contours and ellipses are color-coded by velocity. Thin gray lines show the boundaries of the WAVES coverage.\label{fig11}}
\end{figure*}

\citet{Kent2009} propose that a possible scenario for the origin of the ALFALFA Virgo 7 complex is stripping from a galaxy, with the nearby galaxies NGC 4445 and NGC 4424 (both at similar redshifts to the complex) being identified as possible candidates for the parent galaxy. Our results above, showing an alignment both spatially and in velocity-space between the complex and the tail of NGC 4522, mean that we must also consider the possibility that this galaxy is associated with the complex. In Figure \ref{fig11} we show the region around the complex, with the nearby galaxies labelled and color-coded by redshift, and with \hi\ contours shown for those detected in WAVES. 

An approximate separation/dispersion timescale for the complex can be calculated from the velocity differences between the clouds, $\Delta V$, and their projected separation, $r$, i.e. $\tau \sim r/\Delta V$, assuming a common starting point. It can be seen from Table 2 that a timescale of around 1 Gyr is consistent with the current dispersion of the clumps in the cloud. It can also be seen that if, alternatively, the complex is a bound system (which seems highly unlikely) it would have a dynamical mass of approximately $10^{11}-10^{12}$ \Msol.

\begin{deluxetable*}{lllll}
\tablecaption{Timescales for the clouds to have reached their current separation from the central C1 cloud and implied dynamical mass of the complex (assuming $\delta V = V_\mathrm{circ}$) if it were bound.\label{table2}}
\tablehead{
Cloud&Velocity difference&Projected distance&Timescale&Implied dynamical\\
&from C1 (km/s)&from C1 (kpc)&(Gyr)& mass ($10^{10}$ \Msol).}
\startdata
C2&$111 \pm 4$&67&0.6&19\\
C2b&$110 \pm 7$&68&0.6&19\\
C3&$42 \pm 3$&88&2.0&3.6\\
C4&$70 \pm 4$&38&0.5&4.3\\
C5&$10 \pm 4$&87&8.5&0.2\\
C5b&$189 \pm 8$&91&0.5&76\\
C6&$42 \pm 5$&41&1.0&17\\
\enddata
\end{deluxetable*}

The ALFALFA Virgo 7 complex is similar in some ways to the cloud HI 1225+01 in the Virgo southern extension \citep{Giovanelli1989,Giovanelli1991}. At their assumed distance of 20 Mpc, HI 1225+01 has a total mass of $4.9 \times 10^9$ \Msol\ with the two clumps to the NE and SW of the cloud having masses of $2.4 \times 10^9$ \Msol\ and $1.3 \times 10^9$ \Msol\ respectively \citep{Giovanelli1991}. Approximately a third of the emission is from diffuse gas rather than from the two clumps. This combination of multiple \hi\ clouds and diffuse emission is reminiscent of the complex here, but there are important differences. The more massive NE clump of HI 1225+01 hosts a low surface brightness dwarf irregular galaxy, J1227+0136 \citep{Djorgovski1990,Impey1990,McMahon1990}, although no counterpart has been found to the SW clump despite repeated searches down to 27--28 mag arcsec$^{-2}$ \citep{Impey1990,Salzer1991,Turner1997,Matsuoka2012}. \citet{Chengalur1995} suggested that the SW clump could be an edge-on disc; it has also been suggested \citep{Turner1997} that it could be a tidal tail from the NE clump although this has not been the subject of detailed modeling.

\subsection{NGC 4445 and NGC 4424}

These two galaxies were suggested by \citet{Kent2009} as possible parent galaxies, but the discovery of much more neutral hydrogen in the complex affects the likelihood of this. They have estimated original \hi\ masses based on their optical diameters and morphological types \citep{Solanes1996} of $1.2_{-0.6}^{+0.9} \times 10^9$ \Msol\ and $1.7_{-0.8}^{+1.4} \times 10^9$ \Msol\ respectively, at our adopted distance for the complex of 16.7 Mpc. We measure an \hi\ mass for the complex of $1.3 \pm 0.1 \times 10^9$ \Msol. \citet{Sorgho2017}, using a combined WSRT and KAT-7 map, measure current \hi\ masses for these galaxies, corrected to our adopted distance, of $0.3 \pm 0.1 \times 10^8$ \Msol\ and $2.4 \pm 0.5 \times 10^8$ \Msol. This means they have lost $1.2_{-0.6}^{+0.9} \times 10^9$ \Msol\ and $1.5_{-0.8}^{+1.4} \times 10^9$ \Msol\ respectively. The gas detected in the complex would thus account for 110 percent of the expected gas lost from NGC 4445 or 85 percent of the expected gas lost from NGC 4424. Even taking the upper ends of these distributions, the gas detected in the complex would account for over 60 percent of the gas lost from NGC 4445  and over 45 percent of the gas lost from NGC 4424. By comparison, most previously known \hi\ streams in Virgo account for 20 percent or less of the expected (not upper end) mass lost from their parent galaxies (Taylor et al. 2019, submitted), while the new streams identified in that paper generally contain less than 10 percent of the mass lost from their parents. The ALFALFA Virgo 7 complex would thus, if it originated in either of these galaxies, contain a uniquely high fraction of the missing gas from its progenitor galaxy for any dark feature.

Of the two, the more obvious candidate is NGC 4445: it has lost virtually all of its original gas mass and it is close by and at a similar velocity to the complex. As the ram pressure necessary to account for the deficiency of NGC 4445 is higher than the pressure estimated at its present location, it is likely to have had its stripping in the past \citep{Koeppen2018}. However, as mentioned above, even if it were unusually gas rich the \hi\ mass we measure in the complex would account for over 60 percent of the gas lost from this galaxy. Evaporation for a free-floating cloud is expected to be at a rate of $1-10$ \Msol\ per year, or around $10^9-10^{10}$ \Msol\ over the approximately 1 Gyr timescale of the cloud (Taylor et al. 2019, submitted), so this accounts for the rest of the gas lost. If NGC 4445 is the parent galaxy of the complex, this implies that we have now discovered virtually all of the gas to be found and that it has been removed from the galaxy remarkably efficiently, with little gas dispersed to low column densities. NGC 4445 has a short tail pointing away from the complex \citep{Sorgho2017} and is not well aligned with the complex, although if the stripping occurred near a pericentric passage and both objects have moved and separated since then this is not a strong objection.

NGC 4424 lies close to NGC 4445 both spatially and in velocity. The amount of gas lost from this galaxy is more consistent with the gas seen in the complex, but it is known to be currently losing gas via ram-pressure stripping, with a long \hi\ tail to the south-east \citep{Chung2007,Sorgho2017}. This creates not only an alignment problem -- the tail points nowhere near the complex -- but also the problem that creating the complex as a structure separate from the current tail would require two discrete stripping episodes. It is hard to envision a scenario in which this could occur, without requiring significant, unexpected sub-structure in Virgo's ICM.

\subsection{NGC 4522}
 
\begin{figure}
\plotone{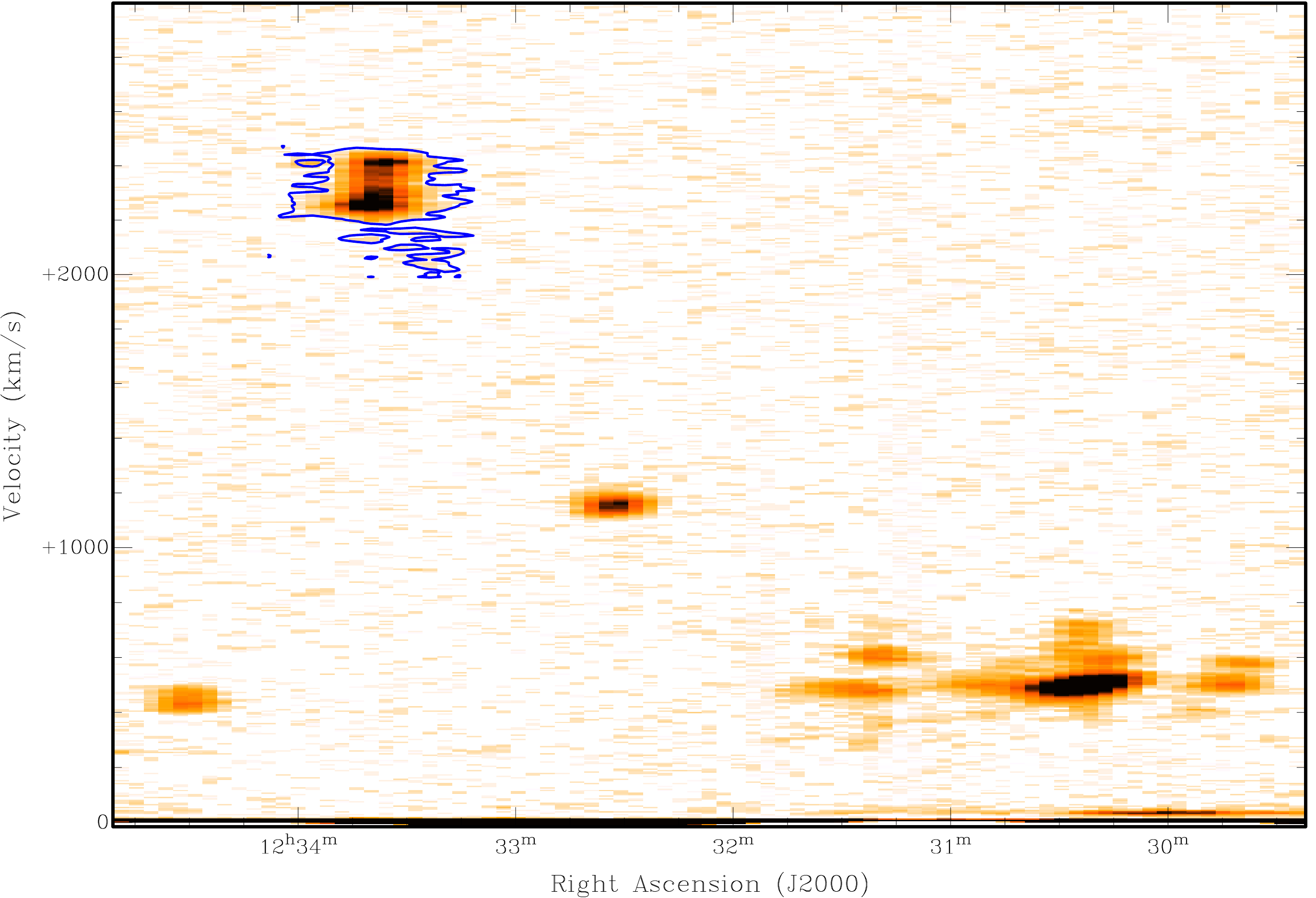}
\caption{Moment 0 map in RA against velocity covering the whole region and clipped at 3$\sigma$. The tail on NGC 4522 is shown by the blue contours (from the position-velocity plot shown in Figure \ref{fig10}, at a flux summed in declination of 5 mJy). The alignment between NGC 4522 and its tail, VCC 1437, and the eastern end of the complex can be easily seen. VCC 1566 is also visible at the east side of the plot.\label{fig12}}
\end{figure}

Could there be a connection between the ALFALFA Virgo 7 complex and NGC 4522? It can be seen from Figures \ref{fig11} and \ref{fig12} that there is an intriguing alignment on the sky between NGC 4522, VCC 1437 and the complex, similar to that expected if the complex were deposited as a stream by NGC 4522. Figure 13 shows that there is also an alignment in velocity space between the current tail and the eastern end of the complex (and VCC 1437), making it very tempting to interpret these as being connected. Based on NGC 4522’s optical radius (SDSS, isophotal) of 120 arcsec and its Sc morphology it is expected to have had \citep[following][]{Solanes1996} an initial mass of $2.5_{-0.9}^{+1.5} \times 10^9$ \Msol, while our measurement of $7.9 \pm 0.5$ \Jykms\ gives an \hi\ mass (at 16.7 Mpc) of $5.2 \pm 0.3 \times 10^8$ \Msol, for a loss of $1.5_{-0.9}^{+1.5} \times 10^9$ \Msol, which is consistent with the \hi\ mass needed to form the complex and gives a deficiency of $0.7 \pm 0.2$.

However, NGC 4522's recessional velocity of 2330 \kms, over 1800 \kms\ higher than the complex, throws up two challenges: Firstly, the extreme velocity difference between the complex and the galaxy makes any connection between the two appear unlikely (\citealt{Kent2009} do not consider NGC 4522 except to note that it is close in projection but “in a different velocity regime”). Secondly, the distance from C3 to NGC 4522 is 68\arcmin, while the distance from C5 is 35\arcmin. If the complex had been lain down as NGC 4522 moved through the cluster, we might expect the velocity C5 to lie about half way between that of C3 and NGC 4522, i.e. about 900 \kms\ higher. But C5 actually lies at a velocity $52 \pm 4$ \kms\ {\it lower} than that of C3. 

Although NGC 4522 lies at a significantly different velocity, it has already been posited \citep{Kenney2004,Vollmer2004,Vollmer2006} to be moving at a high velocity (3500--4000 \kms) relative to the local ICM, either because it is not bound to the cluster or because the ICM itself is in motion due to the infall of the M49 group. The tail to lower velocities revealed by our observations indicates that it is certainly moving at a significantly higher recessional velocity than its local ICM. Gas removed from NGC 4522 would be expected to eventually come to rest at the velocity of the ICM, which could potentially explain the relatively flat velocity profile of the complex and its separation in velocity space from NGC 4522.

The simulations of \citet{Vollmer2006} find a best fit for an inclination of the galactic disc to the ICM wind of 60$\degr$, which gives a maximum line-of-sight component of the ICM wind of around 0.33. For the local ICM to be at the 500 \kms\ velocity of the complex would require this component to be around 0.46 for the highest relative velocity considered (4000 \kms). This would be possible for their 45$\degr$ inclination simulation, which has a maximum line-of-sight component of around 0.56, but this simulation does not reproduce the observed extra-planar gas. It thus seems unlikely that the local ICM line-of-sight velocity in the region could be low enough to explain the  complex as coming from NGC 4522. 

In addition to this, and similarly to NGC 4424, NGC 4522 is currently being stripped and would thus also require two discrete stripping episodes. Putting all of these together, it seems unlikely that NGC 4522 is the parent galaxy for the complex.
 
\subsection{Interactions}

An alternative mechanism to ram-pressure stripping for removing gas from galaxies is tidal interactions, which are thought to be responsible for some of the other dark clouds identified in the Virgo cluster. For example, the VIRGOHI 21 cloud, which is most likely to have been caused by a fairly extreme hyperbolic interaction \citep{Bekki2005,Duc2008,Taylor2017}, has a mass of $2.2 \times 10^8$ \Msol\ \citep{Minchin2005} with a total mass of the cloud and the stream connecting it to NGC 4254 of $4.3 \times 10^8$ \Msol\ \citep{Haynes2007}. NGC 4254 itself, meanwhile, has an \hi\ mass of $4.3 \times 10^9$ \Msol\ \citep{Giovanelli2007}. The event that created VIRGOHI 21 would thus appear to have removed around 5 percent of the mass of the parent galaxy into the compact cloud and a further 5 percent into the stream. Simulations by \citet{Taylor2017} show that, in multiple runs, the parent galaxy typically retains 90--95 percent of its original gas content.

However, unlike ram pressure stripping which only affects the gas in a galaxy, tidal interactions have, in principle, equally strong effects on gas and stars. In practice, the gas disks of galaxies are usually very different from the stellar disks: the gas tends to be more extended than the stars but its density drops exponentially beyond the edge of the stellar disk \citep[e.g.][]{Bigiel2012,Broeils1997}. This means that tidal encounters can disrupt the outermost gas without much affecting the stars, but since the density of this outer gas is low, such disruptions cannot remove significant amounts of gas. The only way for a tidal encounter to remove a high fraction of a galaxy's gas content is if it is also strong enough to affect the inner, denser regions, which would necessarily also disrupte the stellar disk. \citet{Bekki2005}, for example, found that in their simulations \citep[with a model galaxy which followed][in having a gas disk twice the size of the stellar disk]{Broeils1994} the mass fraction of stars within their isolated clouds was 14 to 57 percent. Similarly, \citet{Taylor2017} found that "there is not so much difference in the typical fraction of gas and stars that are stripped", with median fractions remaining in the disk of the parent galaxy of 92 percent for the gas and 95 percent for the stars. It does not seem likely that a tidal interaction with another galaxy could remove most of the gas in a galaxy while leaving that galaxy intact optically, thus a galaxy the size of NGC 4445 or NGC 4424 could not be the parent galaxy in a tidal interaction scenario.

A more plausible alternative, in the abstract, might be to scale up the parent galaxy by an order of magnitude so the removal of $2-3 \times 10^9$ \Msol\ or more is only a relatively small percentage of the total \hi\ mass. This would, however, require a parent galaxy with a few times $10^{10}$ \Msol\ of \hi\ originally \citep[90--95 percent of which would be retained;][]{Taylor2017}, and a similarly massive interactor to pull the gas out -- such galaxies do exist but are rare, and there are no candidates in this part of the Virgo cluster. We would expect that, as with VIRGOHI 21 and the simulations it inspired, at least the parent galaxy would still be relatively nearby, even if a hyperbolic interactor was now too distant to be identified. It thus seems unlikely that interactions can provide a plausible mechanism for the creation of the complex.
 
 \section{Conclusions}

NGC 4522 can be ruled out as a likely source galaxy for the ALFALFA Virgo 7 complex. The galaxy, with its tail pointing towards the eastern end of the complex both spatially and in velocity, is the only large nearby galaxy with any hint of a connection to the complex, but its large velocity separation makes this association dubious at best. Earlier simulations \citep[by][]{Vollmer2006} that might have allowed the gas in the complex to have come to rest with respect to the local ICM do not successfully reproduce the extra-planar gas seen in this galaxy, and their simulations that do reproduce that gas place the recessional velocity of the local ICM near NGC 4522 significantly higher than that of the complex. 

NGC 4445 is a possible source galaxy for the complex, while NGC 4424 is a less likely candidate. Our sensitive 21-cm map of the complex reveals a large reservoir of gas that was not detected by ALFALFA. This gas is found in  low velocity clouds associated with \citet{Kent2009}’s C2 and C5 clouds, in a sixth cloud, termed C6, joined to the main body of the complex, and as low column-density gas between the clouds. Its detection raises the \hi\ mass of the complex from $0.5 \times 10^9$ \Msol\ to $1.3 \times 10^9$ \Msol; this counts against either NGC 4445 or NGC 4424 being a likely source galaxy for the complex. It is notable that NGC 4424 is currently being stripped, while NGC 4445 has lost much of its gas at some point in the past but is not currently thought to be undergoing stripping; It is possible that a single event in the past could have been responsible for this gas loss, from NGC 4445, potentially giving rise to the complex.

For the HI 1225+01 complex, \citet{Matsuoka2012} conclude that its formation process is still not well understood. We must similarly conclude that the origin of the complex remains without a satisfactory solution -- none of the obvious mechanisms normally invoked can explain the presence of an isolated $\sim 10^9$ \Msol\ neutral hydrogen complex in this location in the Virgo cluster. Our WAVES observations will cover the regions to the north and west of the complex over the next couple of years, which will allow us to identify any other gas clouds that might be associated with this complex and potentially give further clues towards its formation. As it stands, the least-worst candidate for the source galaxy remains NGC 4445, but this appears to require it to have been unusually gas-rich prior to the formation of the complex.

 \acknowledgments
 
The Arecibo Observatory was operated at the time of these observation by SRI International under a cooperative agreement with the National Science Foundation (AST-1100968), and in alliance with Ana G. M\'{e}ndez-Universidad Metropolitana, and the Universities Space Research Association. The SOFIA Science Center is operated by the Universities Space Research Association under NASA contract NNA17BF53C. R. Taylor was supported by the Czech Academy of Sciences GA\v CR grant CSF 19-18647S. We thank the anonymous referee for comments that have improved this paper.
 
 \facility{Arecibo Observatory (ALFA)}
 
 \software{Karma \citep{Gooch1996}, MIRIAD \citep{Sault1995}}

 \end{document}